\documentclass{aa}
\usepackage[dvips]{graphicx}
\usepackage{psfig}
\def\AJ{AJ}
\def\ApJ{ApJ}
\def\AA{A\&A}
\def\AAS{A\&AS}

\def\ApJS{ApJS}

\def\MNRAS{MNRAS}

\def\gv{\object{G29.96$-$0.02}}
\def\gt{\object{G31.41$+$0.31}}

\def\HII{\ion{H}{ii}} 
\def\hco{HCO$^+$}
\def\sio{SiO}
\def\hcotr{HCO$^+$(1$-$0)}
\def\siotr{SiO(2$-$1)~v=0}
\def\amm{NH$_{3}$}
\def\ammtr{NH$_3$(4,4)}
\def\methcy{CH$_{3}$CN}
\def\methcytr{CH$_{3}$CN(6$-$5)}

\def\hc5n{HC$_5$N}
\def\wat{H$_{2}$O}
\def\kms{$\rm km\,s^{-1}$}
\def\meth{CH$_{3}$OH}

\begin{document}

%\thesaurus{09(09.03.1; 09.08.1 \gv; 09.08.1 \gt; 09.09.1; 09.13.2: 13.19.3)}
\title{The kinematics of molecular clumps surrounding hot cores in 
\gv\ and \gt}
\author{C. Maxia\inst{1,2}
\and L. Testi\inst{3}
\and R. Cesaroni\inst{3}
\and C.M. Walmsley\inst{3}
}

\offprints{L. Testi, ltesti@arcetri.astro.it}
\institute{
Osservatorio Astronomico di Cagliari, Str. 54, Loc. Poggio dei Pini,
I-09012 Capoterra (CA), Italy
\and
Dipartimento di Scienze Fisiche, Universit\`a di Cagliari, Cittadella
Universitaria, I-09042 Monserrato (CA), Italy
\and
Osservatorio Astrofisico di Arcetri, Largo E. Fermi 5, I-50125 Firenze, Italy}
\date{Received 29 December 2000/Accepted 15 February 2001}

\authorrunning{C. Maxia et al.}
\titlerunning{The hot cores in \gv\ and \gt}

\abstract{
We present high angular resolution interferometric
observations of 
the 3 and 1.3~mm continuum emission, and HCO$^+$(1--0)
and SiO(2--1)~v=0  lines,
obtained with the Owens Valley Radio Observatory millimeter-wave array,
toward two hot cores (HCs) associated with two well known ultracompact
(UC) \HII\ regions: \gv\ and \gt. These HCs 
are believed to host young forming massive stars which have been suggested 
to be surrounded by massive rotating accretion disks.
The aim of these new observations
is to study the structure and kinematics of the molecular clumps surrounding 
the HCs and nearby UC~\HII\ regions at moderately high angular resolution.
Our observations reveal that the clumps within which the
HCs and UC~\HII\ regions are embedded have 
a complex kinematical structure. The total mass
of the clumps is estimated to be in the range 1000--3000~$M_\odot$,
consistent with
previous findings.
Our observations also show compelling evidence that the clump in
\gv\ is contracting onto the HC position, 
suggesting that the accretion
process onto the massive young stellar object embedded in the HC is still
ongoing. In these objects the kinematical structure that we observe is also
compatible with 
the presence of a massive rotating disk within the HC,
even though we cannot prove this suggestion with our data.
The case of \gt\ is more complicated, and our data, although
consistent with the presence of an inner disk and an infalling envelope around
it, do not have the required spatial resolution to resolve the different structures.
\keywords{ISM: clouds -- \HII\ regions -- ISM: individual objects:
          \object{G29.96$-$0.02}, \object{G31.41$+$0.31} -- ISM: molecules -- Radio lines: ISM}
}

\maketitle
                          
\section{Introduction}

The importance of the study of high-mass star formation derives from the
close interaction between such stars and the interstellar medium: they affect
the ISM evolution and, consequently, the galactic evolution in general. In
spite of this, until the last few years the study of star formation
was mostly concentrated on low-mass stars, because of the difficulties
in investigating young massive stars, less numerous, more distant,
more embedded and with much shorter evolutionary timescales.
The search for the sites where the
first phases of high-mass star formation occurs has been increasing recently,
because of the advent of aperture synthesis mm-interferometers
(Kurtz et al.~\cite{kurtz00}).

High-mass stars ignite hydrogen burning and reach the main sequence while
still accreting matter (Palla \& Stahler~\cite{PS93}). Thus young massive
stars are still deeply embedded in a dense gas and dust envelope which
impairs the detection at optical and near infrared wavelengths. Due to
the large luminosity and Lyman photon emission, a newly born massive star
heats the dust and ionises the gas in the surrounding envelope and can be
detected in the far-infrared (FIR) as a luminous IRAS source and in the
centimetric radio continuum as an ultracompact (UC) \HII\ region (Wood \&
Churchwell~\cite{WC89a}; \cite{WC89b}). However, these are signposts
of already-formed massive stars; in order to study the more elusive
evolutionary phase of main accretion one has to devise suitable target
selection procedures. Several different approaches have been explored
to obtain lists of candidate massive accreting protostars
(see the review by Kurtz et al.~\cite{kurtz00}): high-luminosity
IRAS sources with peculiar colours (e.g. Molinari et al.~\cite{Mea96};
\cite{Mea98}), H$_2$O masers
without associated radio continuum emission (Tofani et al.~\cite{Tea95};
Cesaroni et al.~\cite{CFW99}),
compact molecular cores close to UC \HII\ regions (Cesaroni et al.~\cite{Cea91};
\cite{cesa94}). The latter approach is supported by the observational result
that massive stars are usually found in rich clusters and associations: it
is thus likely to find a forming massive star lying close to a young one. 
In this way Cesaroni
et al.~(\cite{cesa94}) identified compact molecular cores close to,
but generally not coincident with, the target UC \HII\ regions.
These are characterized by bolometric luminosities exceeding $10^4$~$L_\odot$,
small diameters ($\leq 0.1$~pc)  
and kinetic temperatures over
50~K, and were then called hot cores (HCs). Due to their high luminosity
and temperature, HCs have been suggested to contain very young massive
(proto-)stars.  In fact, a few of them contain UC \HII\ regions (e.g.
\object{G10.47+0.03}, see Cesaroni et al.~\cite{cesa98}), which witness
the existence of embedded early type stars.
Moreover, two well known HCs, \object{IRAS~20126$+$4104}
(Cesaroni et al.~\cite{Cea97}; \cite{Cea99}) and \object{G9.62+0.19--F} (Hofner et
al.~\cite{Hea96}; Testi et al.~\cite{Tea98}; \cite{Tea00}), show evidence
of an embedded massive young stellar object in a phase preceding the
formation of a classical UC \HII.

Two of the best studied classical HCs from the list of Cesaroni et
al.~(\cite{cesa94}) are those close to the UC \HII\ regions \gv\ and \gt.
In the former region the molecular core lies in front of the cometary
UC \HII\ region, approximately 2$\farcs$6 west of the 1.3 cm continuum
peak. In \gt, faint \amm\ emission is detected toward the extended halo
around the UC \HII\ region, while the \amm\ HC peak is $\sim$4\arcsec\
to the southwest of the centimetric radio continuum peak. High angular
resolution ($\sim$0.4\arcsec) Very Large Array (VLA)
observations of the \ammtr\ inversion
transition (Cesaroni et al.~\cite{cesa98}) showed resolved emission from
both HCs, with a deconvolved size of $\sim$1\farcs4 in \gv\ and
$\sim$1\farcs8 in \gt. Both cores appear slightly elongated: \gv\ in the
E-W direction and \gt\ along a SW-NE axis. A velocity gradient
of approximately 2-3~km~s$^{-1}$~arcsec$^{-1}$ over 3\arcsec\ is found
along such directions.  Similar correlations are found in the \methcytr\
transition for both \gv\ and \gt, although the velocity gradients in \methcy\
($\sim$15~km~s$^{-1}$~arcsec$^{-1}$) are larger than those observed
in \amm\ (Cesaroni et al. \cite{cesa31.94}; Hofner et al. in prep.).
Also, in both sources the derived
kinetic temperature shows a peak toward the center.
All these features have been interpreted as evidence
of disk-like structures rotating about a hot, massive central object:
the (proto-)star.

The presence of a deeply embedded massive \mbox{(proto-)}star is also indirectly
suggested by the presence of several other signposts, such as
\wat\ and H$_2$CO, \meth\ masers toward the hot ammonia core peak of \gv\
(Hofner \& Churchwell \cite{hofner96}; Pratap et al \cite{pratap94};
Walsh et al. \cite{walsh98}) and \wat\ masers toward \gt\ (Gaume \&
Mutel~\cite{GM87}).

Recently, Pratap et al.~(\cite{pratap99}) investigated the molecular gas
surrounding \gv\ using the BIMA millimeter wave interferometer with moderate
angular resolution ($\sim$10\arcsec). They found $^{13}$CO and high-density
tracer (CS and CH$_3$CN) emission; while the former molecule is found in an
extended cloud surrounding both the HC and the UC~\HII, the latter peak
toward the HC position.
Hence, \methcy\ and \amm\ emission trace the innermost and warmer molecular gas
within the HC itself, while $^{13}$CO traces the large-scale molecular
cloud in which the HC, the UC \HII\ region and the surrounding near infrared
young stellar cluster are embedded.
Pratap et al.~(\cite{pratap99}) also detected a
compact core of \meth\ emission, $\sim$5\arcsec\ south west of the HC. This
second core does not show the typical features of HCs (compact \amm\ and
\methcy\ emission) and Pratap et al. suggest that it may be associated with a
molecular
outflow from the HC exciting source or with a low-mass star forming outside it.

In order to study the relationship between the HC and the UC \HII\ region, and
the kinematics of the gaseous clump in which these are embedded, 
we decided to obtain high angular resolution observations of 
the HCO$^+$(1--0) and \siotr\ lines, as already done in similar regions
by other authors (Acord, Walmsley \& Churchwell~\cite{Aea97}; Cesaroni et
al.~\cite{Cea97}; Wilner et al.~\cite{Wea96}).
These tracers are expected to provide information
on the innermost part of the cloud and the kinematics of the molecular gas
at the interface between the HC and the UC \HII\ region. The main goal of the
observations was thus to relate the observed velocity field within the HC
to the kinematics of the molecular gas in the inner portions of the cloud.
Additionally, the \siotr\ transition, which can be enhanced
in shocked regions, was expected to clarify the \meth\ observations
of Pratap et al.~(\cite{pratap99}).

Throughout this paper we will assume a distance of 6~kpc for \gv,
based on the recent estimate of Pratap et al.~(\cite{pratap99}),
and 7.9~kpc for \gt, based on the kinematic estimate of Churchwell et
al.~(\cite{churchwell90}).

We present our new millimeter interferometric observations and data reduction
in Sect.~\ref{sobs}. 
In Sect.~\ref{sres} the observational results are presented for the millimeter
continuum and molecular lines separately.  
We discuss the implications of our results on the structure of the
two regions and the nature of HCs in Sect.~\ref{sdisc}.
In Sect.~\ref{sconc} we summarize the main results of our study.

\section{Observations and data reduction}
\label{sobs}

High resolution interferometric observations were performed
in the period October-December 1997 using
the Owens Valley Radio Observatory (OVRO) millimeter-wave array,
located near Big Pine, California. The array comprises six 10.4~m dishes
that were employed in three different configurations yielding baselines
in the range 15-220~m. The phase centers used were:
$\alpha_{1950}=18^{\rm h}43^{\rm m}27\fs1$, $\delta_{1950}=-02\degr 42\arcmin 36\farcs4$
for \gv\ and
$\alpha_{1950}=18^{\rm h}44^{\rm m}59\fs1$, $\delta_{1950}=-01\degr 16\arcmin 07\farcs3$
for \gt. The half-power beamwidth of the OVRO antennas is
$\sim$84\arcsec\ at 88~GHz, and $\sim$32\arcsec\ at 233~GHz.
Each antenna is equipped with cryogenically cooled SIS receivers that
offered typical system temperatures of $\sim$300~K at 88~GHz, and
$\sim$1200~K at 233~GHz. The 3.4~mm and 1.3~mm continuum observations
employed an analog correlator for a total bandwidth, at each of the
two frequencies, of $\sim$2~GHz. The flexible digital correlator
was configured to observe simultaneously the \hcotr\ and \siotr\
transitions at 89.2 and 86.8~GHz respectively. For both molecules
we used a $\sim$105~km/s wide band with $\sim$1.7~km/s resolution.
The spectrometer modules were centered at $\sim$98~km/s, close
to the systemic velocity of both sources.

Complex gain calibration was ensured by frequent observations of 1741$-$038.
The typical observing cycle alternated scans on \gv, \gt\ and 1741$-$038
with a period of $\sim$30~min. 3C273 was used as passband calibrator
and Uranus and/or Neptune were used to set the flux scale. The expected
flux calibration uncertainty is $\sim$20\%.
Data calibration and editing has been performed using the MMA
software package (Scoville et al.~\cite{Sea93}). At 230~GHz data
could be salvaged only for a small subset of the observations,
due to the poor phase stability and relatively high water vapour
content of the atmosphere during most of the observations.
This resulted in a poor ($u,v$) coverage of the final 233~GHz
dataset, a final noise higher than normally expected and a much reduced
sensitivity to the extended structures. The calibrated
($u,v$) datasets were exported to the NRAO--AIPS package
for imaging and analysis. All the maps presented in this paper
were obtained with the AIPS IMAGR task using natural weighting
of the ($u,v$) data.

In the continuum maps, the final synthesised half power beam width (HPBW)
is $5 \farcs 9 \times 3 \farcs 7$, with P.A.$\sim -8$, at
88~GHz, and $4\farcs9 \times 2\farcs 7$, with P.A.$\sim -3.5$, at 233~GHz, 
for both sources.
The rms level is 10 mJy/beam at 88~GHz for both sources, 40 mJy/beam
for \gv\ and 160 mJy/beam for \gt\ at 233~GHz.  
In the line cubes we obtained 
$5 \farcs 9 \times 3 \farcs 7$ HPBW and 50 mJy/beam rms level for
all lines. The P.A. is $-10.5$ for the \hco\ and $-16.1$ for the
\sio\ for \gv; $-9.2$ for the \hco\ and $-14.9$ for the \sio
line for \gt.
The conversion factor from flux density  to brightness temperature in the
synthesised beam is $\sim$10.4 K (Jy/beam)$^{-1}$ at 88~GHz and
$\sim$2.43 K (Jy/beam)$^{-1}$ at 233~GHz.

\subsection{Continuum maps}
\label{sobs_cont}

\begin{figure}
\centerline{\psfig{figure=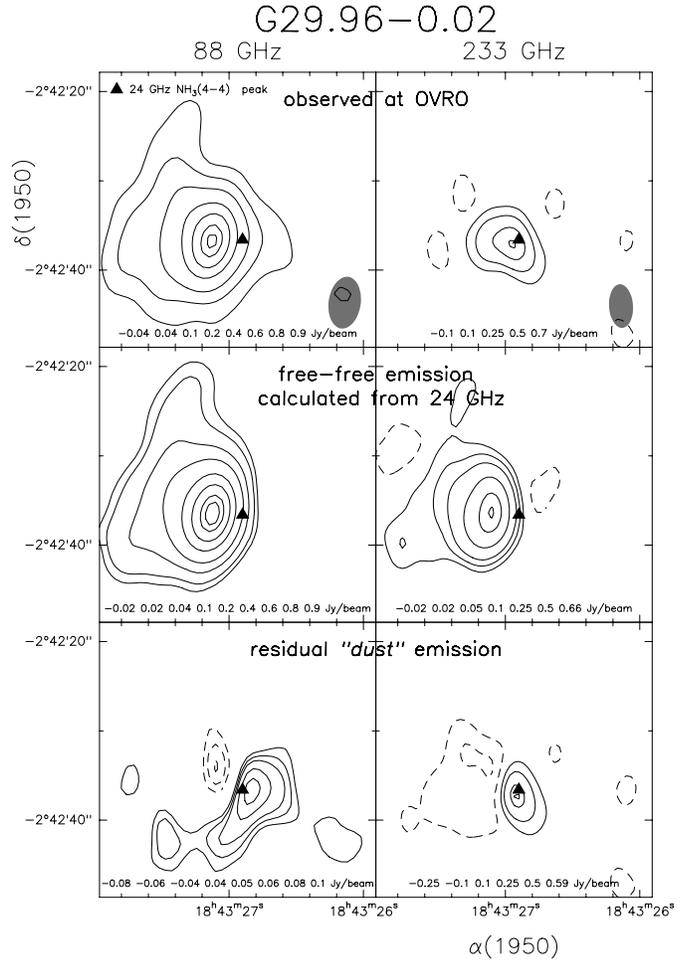,width=8.8cm}}
\caption{
Top panels: 88 and 233~GHz millimeter continuum maps observed at
OVRO towards \gv. Central panels: expected free-free contribution computed
from the 24~GHz VLA data of Cesaroni et al.~(\cite{cesa94}; \cite{cesa98})
using the same ($u,v$) range and restoring beam as in the top panels.
Bottom panels: residual ``dust emission'' maps after subtracting the central
panels from the top panels maps. The filled triangle marks the ammonia
HC peak position.
The beamsize at each frequency is shown on the top panels.}
\label{g29_cont}
\end{figure}
\begin{figure}
\centerline{\psfig{figure=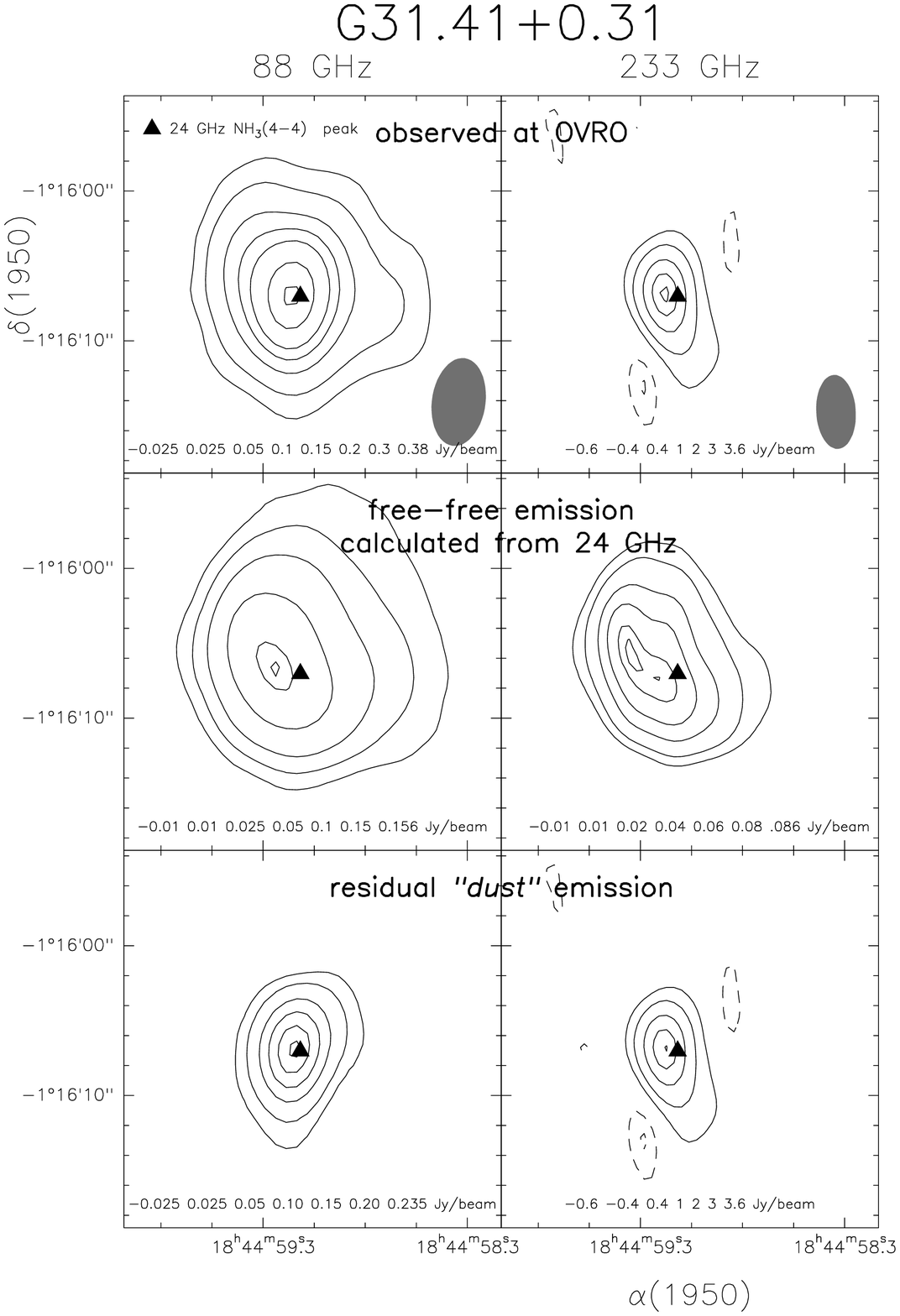,width=8.8cm}}
\caption{As in Fig.~\ref{g29_cont}, for \gt.}
\label{g31_cont}
\end{figure}

Our millimeter continuum maps are sensitive to both the free-free emission
from the ionised gas inside the UC \HII\ regions and the thermal dust
emission. The latter is mainly confined to the HC region
(Olmi et al.~\cite{olmi96}; Patrap et al.~\cite{pratap99}).
In order to disentangle the HC continuum emission from that of the
UC \HII\ region, we need to estimate the contribution of the free-free emission
at millimeter wavelengths and subtract it from our continuum maps.
To this purpose we have used the 24~GHz VLA observations of Cesaroni
et al.~(\cite{cesa94}; \cite{cesa98}). We re-imaged the calibrated
($u,v$) data using the same ($u,v$) range and restoring beam
as the 88 and 233~GHz OVRO observations. These maps were then scaled
assuming optically thin free-free emission ($F_\nu\propto\nu^{-0.1}$).
In Figs.~\ref{g29_cont} and~\ref{g31_cont},
we show for each source and each frequency the millimeter continuum maps
as observed at OVRO (top panels), the free-free emission maps obtained
by extrapolation of the VLA data (central panels),
and the residual ``dust'' emission maps
after removal of the free-free emission from the millimeter observations
(bottom panels).

Figures~\ref{g29_cont} and~\ref{g31_cont} show that the procedure described
above allowed us to compute a fairly accurate estimated map of the free-free
emission contribution at millimeter wavelengths. The subtraction of the
extended UC \HII\ is successful in all cases except for the 233~GHz observations
of \gv. In this case, the poor ($u,v$) coverage of the millimeter map
does not allow a proper reconstruction of the extended emission, and
even if the free-free map was calculated using the same ($u,v$) range as
the millimeter map, the much finer VLA sampling of the ($u,v$) plane allowed
us to properly map part of the extended emission. When the subtraction of
the calculated map is performed, the free-free emission from the UC \HII\
is oversubtracted, producing a large negative hole at the position
of the UC \HII. On the other hand, also in this case, the position of the
excess ``dust'' emission at 233~GHz is coincident with the \amm\ core and
the 88~GHz ``dust'' peak.
Toward \gt\ the free-free contribution to the millimeter emission is
lower: at 88 GHz, the observed continuum peak is
twice the value calculated from the 1.3~cm map, whereas at 233~GHz the maximum
of the expected free-free emission is
less than the 1~$\sigma$ noise level in the observed continuum map.
In this case, the observed flux at 233~GHz is essentially thermal dust emission.
However, as for \gv,
the total continuum flux at 233~GHz measured in our interferometric maps is
less (a factor of $\sim 3$ for \gt)
than what expected from previous single dish observations (see
Fig.~\ref{sed}),
probably due to the spatial filtering of extended emission
(see also Sect.~\ref{sres_cont}).

\subsection{Line maps}

Continuum subtraction from the line
data cubes was performed both in the ($u,v$) plane using the AIPS task
UVLIN, selecting line free channels prior to image deconvolution, and
in the image plane after cleaning, by subtracting the
appropriate
continuum images.
The two techniques offered comparable results.
The analysis presented in this paper
uses the continuum subtracted cubes produced with the second method.

\section{Results}
\label{sres}

\begin{figure*}[th]
%\resizebox{\hsize}{!}{\includegraphics{sed.eps}}
\centerline{\psfig{figure=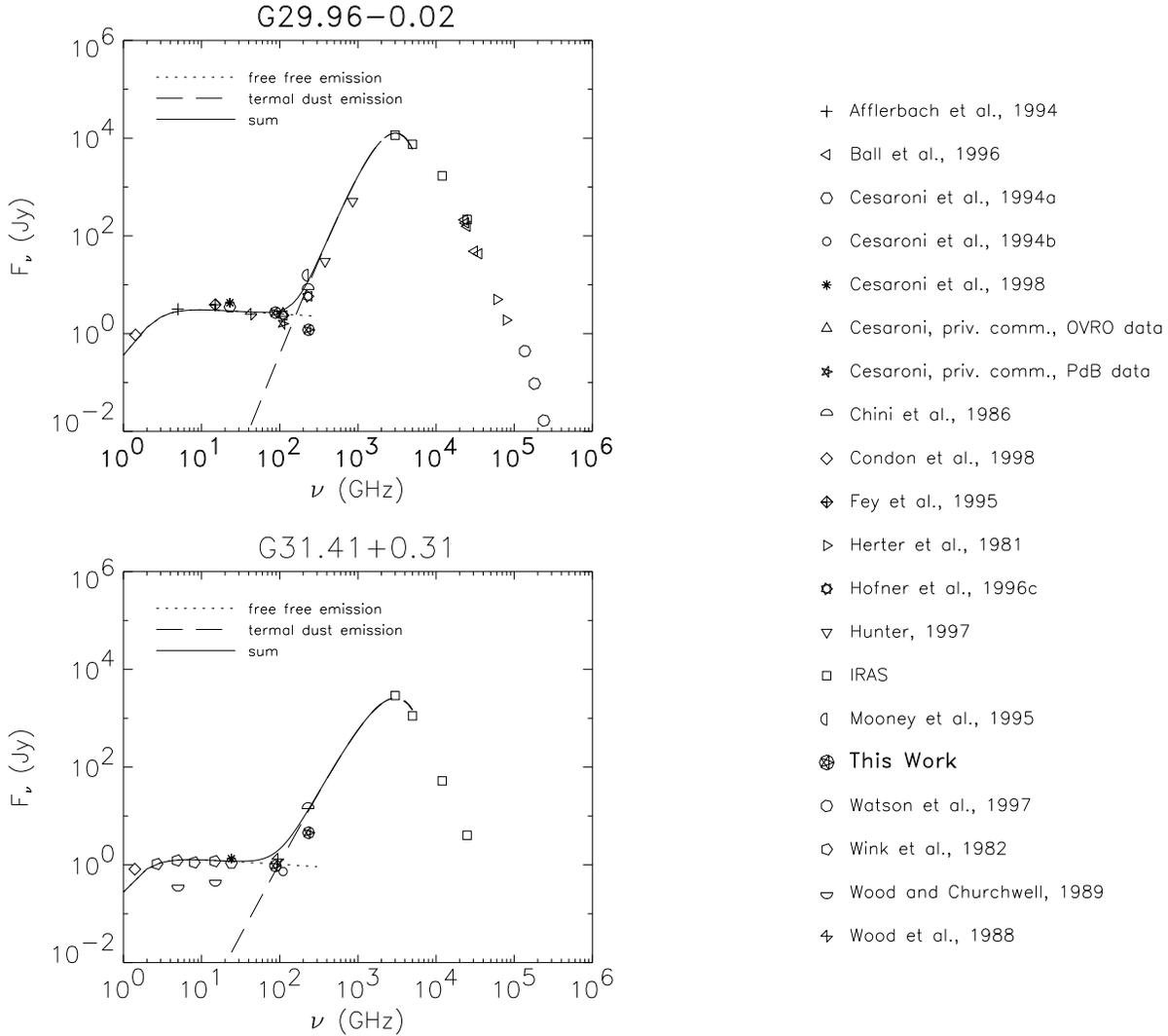,height=14.5cm}}
\caption{Spectral energy distribution for \gv\ and \gt. The solid lines
are fits of the SED assuming a simplified two component model: free-free
emission from a spherical, homogeneous and dust-free \HII\ region (dotted
line), and grey-body thermal dust emission (dashed line). On the right we
indicate the reference for each of the measurement reported in the plots.}
\label{sed}
\end{figure*}

\begin{table}
\caption[]{SED fit parameters} \label{tabfitsed}
\begin{flushleft}
\begin{tabular}{*{7}{c}}
\noalign{\smallskip}
\hline
\noalign{\smallskip}
source& $d$ & $r_{\rm H\,{II}}$ & $T_{\rm H\,{II}}$&$\log N_{\rm Ly}$&$T_{\rm dust}$&$\beta$\\
      &(Kpc)&(pc)         &(K)          &(ph./s)          &(K)  &       \\
\hline
\gv\  &6    &0.5          &6500         &48.90            &30   &2      \\
\gt\  &7.9  &0.5          &8000         &48.75            &37   &1      \\
\noalign{\smallskip}
\hline
\end{tabular}
\end{flushleft}
\end{table}

\subsection{Continuum}
\label{sres_cont}

\begin{table*}
\caption[]{Observed parameters of the continuum (dust and free-free) emission} \label{tabcont}
\begin{flushleft}
\begin{tabular}{*{9}{c}}
\noalign{\smallskip}
\hline
\noalign{\smallskip}
source  &  $\nu$   & $\alpha_{1950}^{\rm peak}$ & $\delta_{1950}^{\rm peak}$ &
  $F_\nu^{\rm peak}$  &      $F_\nu$ & $\Theta_{\rm 3\sigma}$ & $\theta_{\rm 3\sigma}$ &
   $r_{\rm 3\sigma}$ \\
    &(GHz)&&&(Jy/beam)&(Jy)&($^{\prime\prime}$)&($^{\prime\prime}$)&(pc)\\
\noalign{\smallskip}
\hline
\noalign{\smallskip}
\gv & 88&18:43:27.15&$-$2:42:37.1&0.92&2.7 &20 &20 &0.28\\
    &233&18:43:26.93&$-$2:42:36.9&0.71&1.2 &8.1&6.9&0.10\\
\gt & 88&18:44:59.15&$-$1:16:07.3&0.39&0.92&16 &15 &0.28\\
    &233&18:44:59.13&$-$1:16:06.8&3.7 &4.2 &7.1&5.7&0.11\\
\noalign{\smallskip}
\hline
\end{tabular}
\end{flushleft}
\end{table*}

\begin{table*}
\caption[]{Derived parameters of the dust continuum emission from the cores} \label{tabdust}
\begin{flushleft}
\begin{tabular}{*{9}{c}}
\noalign{\smallskip}
\hline
\noalign{\smallskip}
source  &  $\nu$   & $\alpha_{1950}^\mathrm{peak}$ & $\delta_{1950}^{\rm peak}$ &
  $F_\nu^{\rm peak}$  &      $F_\nu$  &$\Theta_{\rm 3\sigma}$ &  $\theta_{\rm 3\sigma}$
  & $r_{\rm 3\sigma}$ \\
    &(GHz)&&&(Jy/beam)&(Jy)&($^{\prime\prime}$)&($^{\prime\prime}$)&(pc)\\
\noalign{\smallskip}
\hline
\noalign{\smallskip}
\gv & 88&18:43:26.82&$-$2:42:36.4&0.11&0.24&11 &9.5&0.14\\
    &233&18:43:26.90&$-$2:42:37.4&0.60&0.56&5.8&4.0&0.06\\
\gt & 88&18:44:59.10&$-$1:16:06.6&0.24&0.28&9.5&7.8&0.15\\
    &233&18:44:59.13&$-$1:16:06.8&3.6 &4.0 &6.9&5.5&0.10\\
\noalign{\smallskip}
\hline
\end{tabular}
\end{flushleft}
\end{table*}

In Fig.~\ref{sed} we show the spectral energy distribution (SED) for \gv\
(top) and \gt\ (bottom). The flux density measurements that we used are a
compilation of single dish and interferometer observations (see references
in Fig.~\ref{sed}). The data are heterogeneous in both angular resolution
and ($u,v$) coverage. The SED is that typical of UC \HII\ regions, see also
Wood \& Churchwell~(\cite{WC89b}), dominated by free-free emission
at centimeter wavelengths
and by thermal dust emission at millimeter and far-infrared 
wavelengths. 

To give an estimate of global parameters, such as the supply rate of Lyman
photons required to mantain the UC \HII\ region ionised and the average
temperature of the dust, we used a simplified two component model:
a spherical and homogeneous \HII\ region, plus a single temperature grey-body
dust component. This simplified model is able to accurately reproduce
the radio to far infrared portion of the SEDs shown in Fig.~\ref{sed}.
In Table~\ref{tabfitsed} we report the values of the fixed parameters of the
models: distance ($d$) and  electron
temperature ($T_{\rm H\,{II}}$) of the  UC \HII\ regions. For \gv, we used the values estimated by
Pratap et al.(\cite{pratap99}); for \gt, the distance is that estimated by Churchwell et
al.~(\cite{churchwell90}), whereas the
temperature is a mean value for a typical \HII\ region.
Also shown in the table are the values of the parameters derived from our model
fits: UC \HII\ radius ($r_{\rm H\,{II}}$), Lyman photons supply rate ($N_{\rm Ly}$), dust temperature ($T_{d}$),
and dust opacity index ($\beta$).

From the SEDs reported in Fig.~\ref{sed}, we can clearly see that our 233~GHz 
observations are well below both the single dish points and the model fits,
expecially in the case of \gv, where the contribution of the UC \HII\ free-free
emission at such high frequency is still a consistent fraction of the
total emission. This is mainly due to the sparse ($u,v$)
coverage of our synthesis at this wavelength, and the inability to properly
image extended structures. Hence, as discussed in Sect.~\ref{sobs_cont}, our
233~GHz observations cannot be used to infer the properties of the
large scale structures in these regions, but offer only an estimate of
the high frequency emission of the HCs themselves.

The values of $T_{\rm dust}$ and $\beta$  are average values for the
dust in the envelope surrounding the HCs, not the appropriate values 
for the HCs since these will only contribute to a small
fraction of the single dish submillimeter and far infrared fluxes.
Nevertheless, the dust temperature that we derive is probably 
a good estimate of the kinetic temperature of the HCO$^+$/SiO molecular clump
surrounding the HC and UC \HII\ region (see Sect.~\ref{sres_lines} below).

In Table~\ref{tabcont}, for each source and at each frequency,
we report the peak position ($\alpha_{1950}^\mathrm{peak}$ and $\delta_{1950}^{\rm peak}$),
the peak ($F_\nu^{\rm peak}$)  and integrated fluxes ($F_\nu$) ,
the observed ($\Theta_{\rm 3\sigma}$) and deconvolved ($\theta_{\rm 3\sigma}$)
angular diameter, and
the corresponding linear radius ($r_{\rm 3\sigma}$)
of the observed continuum emission (dust plus free-free, Figs.~\ref{g29_cont}
and~\ref{g31_cont} top panels).
In Table~\ref{tabdust} the same quantities are 
reported for the estimated dust emission
(Figs.~\ref{g29_cont} and~\ref{g31_cont}
bottom panels). The observed angular diameters
of the sources have been computed as
$\Theta_{\rm 3\sigma}=2\sqrt{A/\pi}$, where $A$ is the area encompassing
the observed source emission inside the 3~$\sigma$ contour level.
The values of $\Theta_{\rm 3\sigma}$ have then been deconvolved
using the expression $\theta_{\rm 3\sigma} =
\sqrt{\Theta_{\rm 3\sigma}^2 - \Theta_{\rm B}^2}$, where $\Theta_{\rm B}$ 
is the synthesized HPBW. The linear radii have been calculated
from $\theta_{\rm 3\sigma}$ using the adopted distances of 6 and 7.9~kpc, for
\gv\ and \gt\ respectively (Pratap et al.~\cite{pratap99}; Churchwell et
al.~\cite{churchwell90}).

\subsection{Line}
\label{sres_lines}

\begin{figure}
\centerline{\psfig{figure=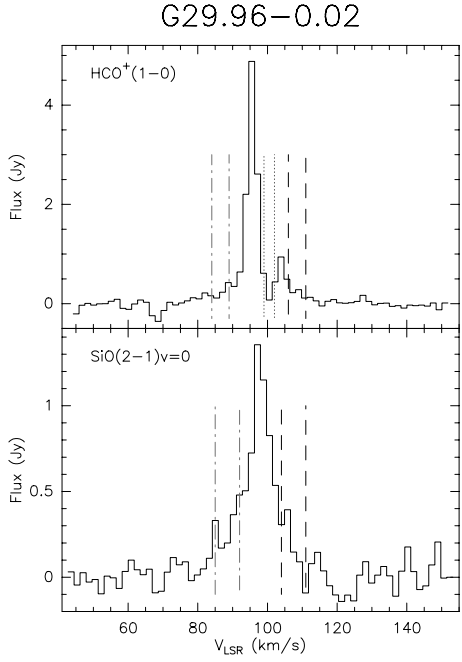,width=6.0cm}}
\caption{\gv\ source integrated spectra of the  \hcotr\ and \siotr\ lines.
The vertical lines indicate the velocity ranges used
for the maps shown in Fig.~\ref{g29_righe}
(see text for details).}
\label{g29_spettri}
\end{figure}

\begin{figure}
\centerline{\psfig{figure=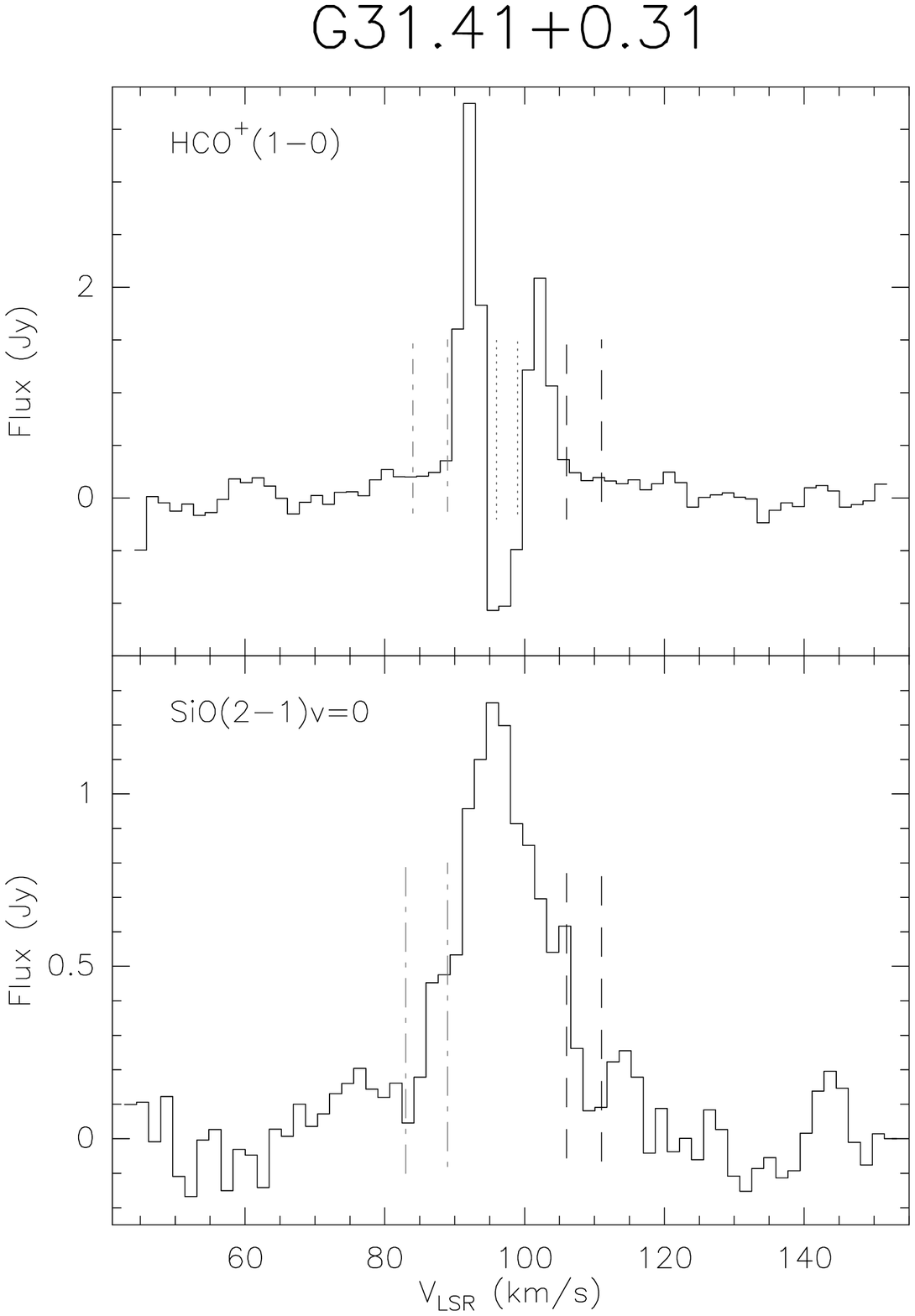,width=6.0cm}}
\caption{As Fig.~\ref{g29_spettri}, for \gt.}
\label{g31_spettri}
\end{figure}

\begin{figure}
\centerline{\psfig{figure=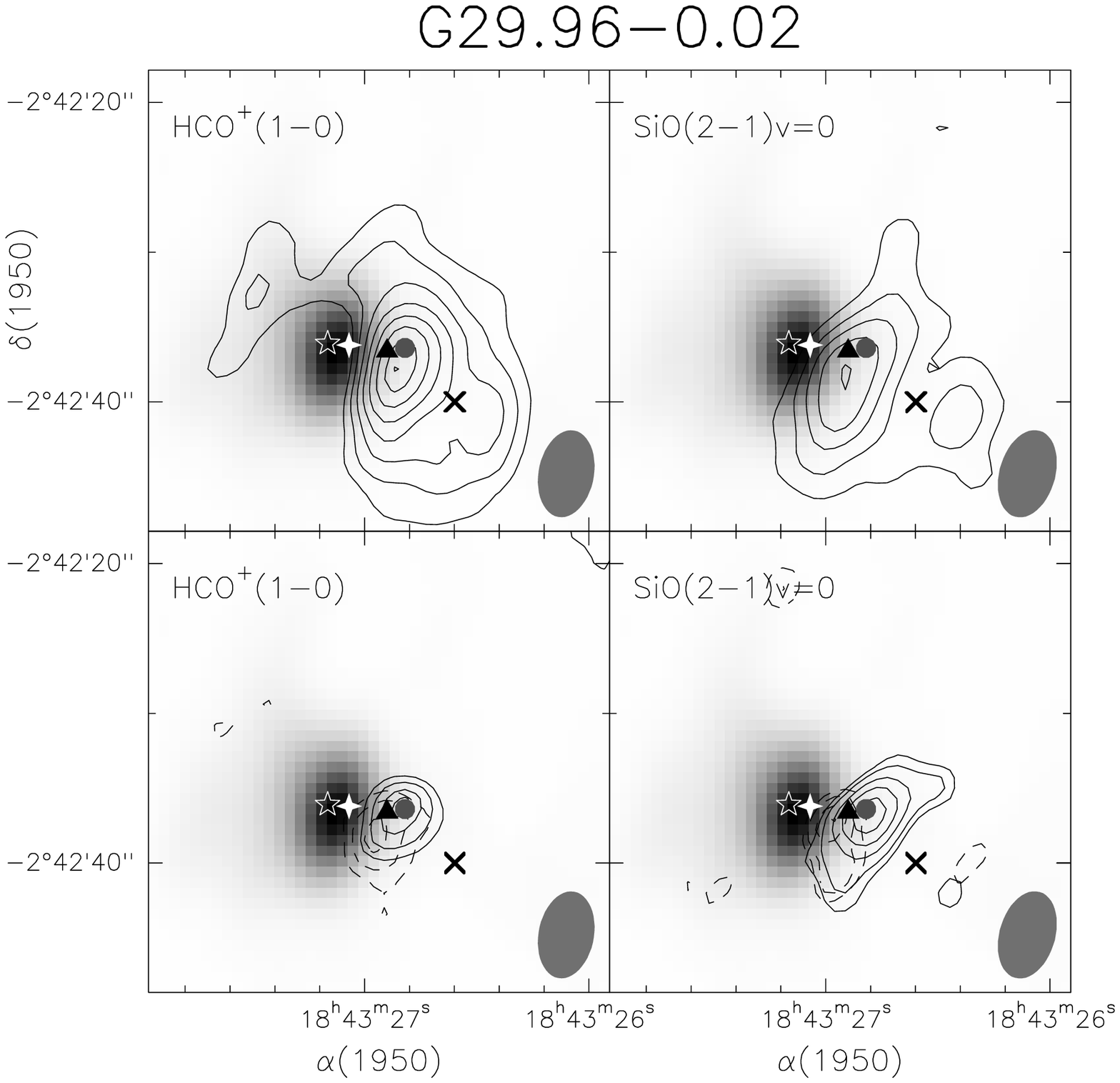,width=8.8cm}}
%\begin{figure*}
%\centerline{\psfig{figure=g29_ali.eps,width=15cm}}
\caption{
Maps toward \gv\  in the \hcotr\ and \siotr\  
emission averaged in velocity over all the line (top)
and over the {blue} (solid contours) and {red} 
(dashed contours) {wings} (bottom),
superimposed to the 88 GHz continuum map (grey scale).
The ranges of velocity used in the maps  are:
for the \hco\ 84--111~km/s for all the line, 84--89~km/s for 
the {blue wing} and 106--111 km/s for the {red wing};
for the \sio\ 85--111~km/s for all the line, 85--92~km/s for 
the {blue wing} and 104--111 km/s for the {red wing}.
The contour levels for the \hco\ line are:
$-0.10$, 0.10, 0.15, 0.20, 0.25, 0.3, 0.35, 0.4, 0.43 Jy/beam;
$-0.09$, 0.09, 0.12, 0.15, 0.17 Jy/beam ({blue})
 and $-0.09$, 0.09, 0.12, 0.15 Jy/beam ({red}).
The contour levels for the \sio\ are:
$-0.05$, 0.05, 0.10, 0.15, 0.20, 0.25 Jy/beam;
$-0.09$, $-0.07$, 0.07, 0.09, 0.12, 0.15, 0.18, 0.20 Jy/beam ({blue})
and $-0.09$, $-0.07$, 0.07, 0.09, 0.12, 0.145 Jy/beam ({red}).
The star marks the position of the UC \HII\ region ionizing star, the circle
is the 88~GHz \emph{dust} peak, the filled diamond the 24~GHz continuum peak,
the triangle the \ammtr\ HC peak, and the cross marks the peak position of the
CH$_3$OH clump detected by Pratap et al.~(\cite{pratap99}).  
}\label{g29_righe}
%\end{figure*}
\end{figure}

\begin{figure}
\centerline{\psfig{figure=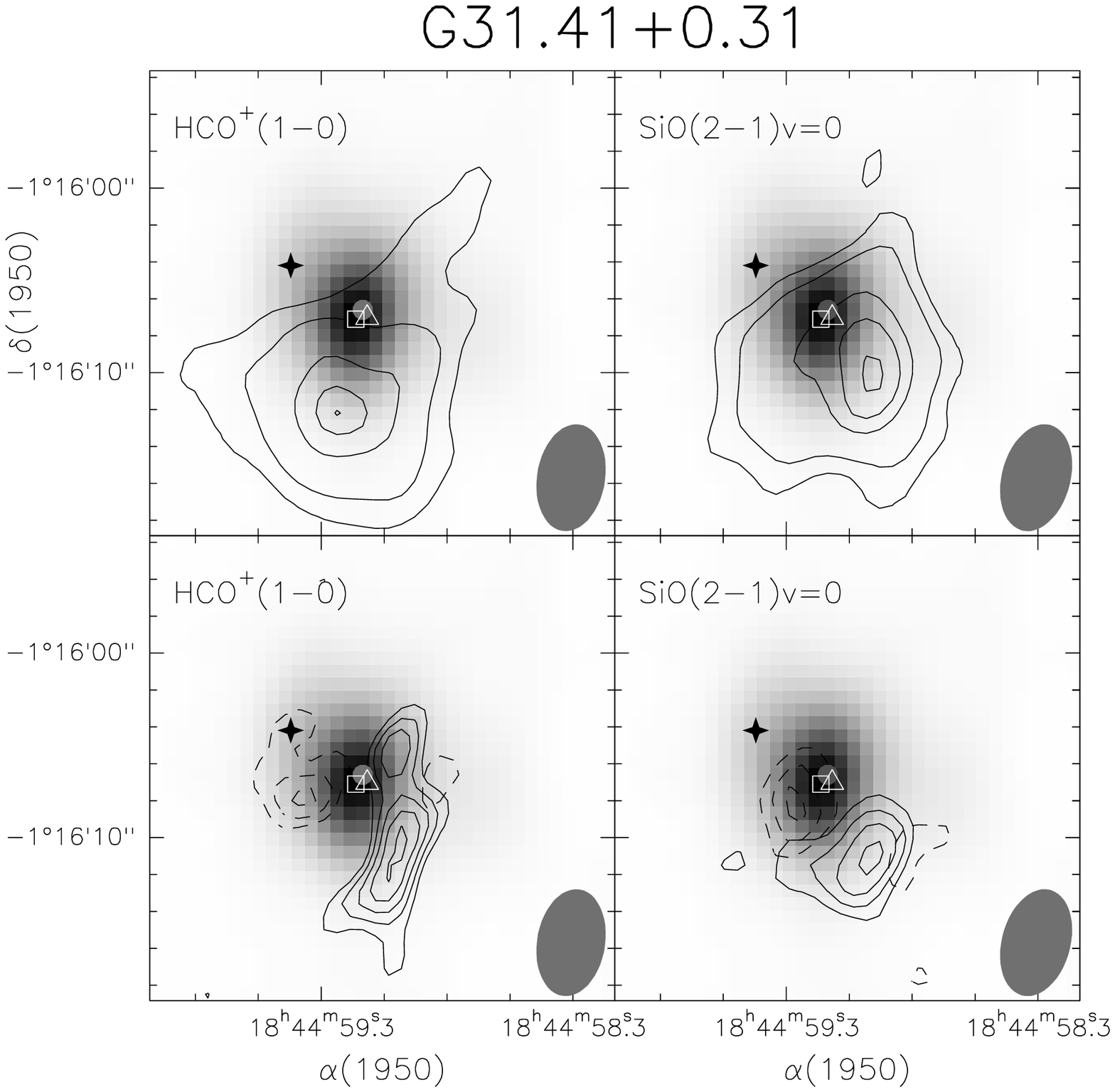,width=8.8cm}}
%\begin{figure*}
%\centerline{\psfig{figure=g31_ali.eps,width=15cm}}
\caption{
Maps toward \gt\ in the \hcotr\ and \siotr\ emission averaged in velocity over
all the line (top)
and over the {blue} (solid contours) and {red} (dashed contours) 
{wings} (bottom),
superimposed to the 88 GHz continuum map (grey scale).
The ranges of velocity used in the maps  are:
for the \hco\ 84--111~km/s for all the line, 84--89~km/s for 
the {blue wing} and 106--111 km/s for the {red wing};
for the \sio\ 83--111~km/s for all the line, 83--89~km/s for 
the {blue wing} and 106--111 km/s for the {red wing}.
The contour levels for the \hco\ line are:
$-0.05$, 0.05, 0.10, 0.20, 0.25, 0.27 Jy/beam;
$-0.08$, 0.08, 0.10, 0.12, 0.14, 0.16 Jy/beam ({blue}) and
$-0.08$, 0.08, 0.10, 0.12 Jy/beam ({red}).
The contour levels for the \sio\ are:
$-0.05$,  0.05,  0.08, 0.12, 0.16, 0.20, 0.24 Jy/beam;
$-0.09$, 0.09, 0.12, 0.15, 0.18 Jy/beam ({blue}) and
$-0.09$, 0.09, 0.12, 0.14 Jy/beam ({red}).
The square marks the position of the 110~GHz continuum peak, the circle is 
the 88~GHz \emph{dust} peak, the filled diamond the 24~GHz continuum peak 
and the triangle the \ammtr\ peak. 
}\label{g31_righe}
%\end{figure*}
\end{figure}

\begin{figure}
%\resizebox{\hsize}{!}{\includegraphics{g29_ass_99_102.eps}}
\centerline{\psfig{figure=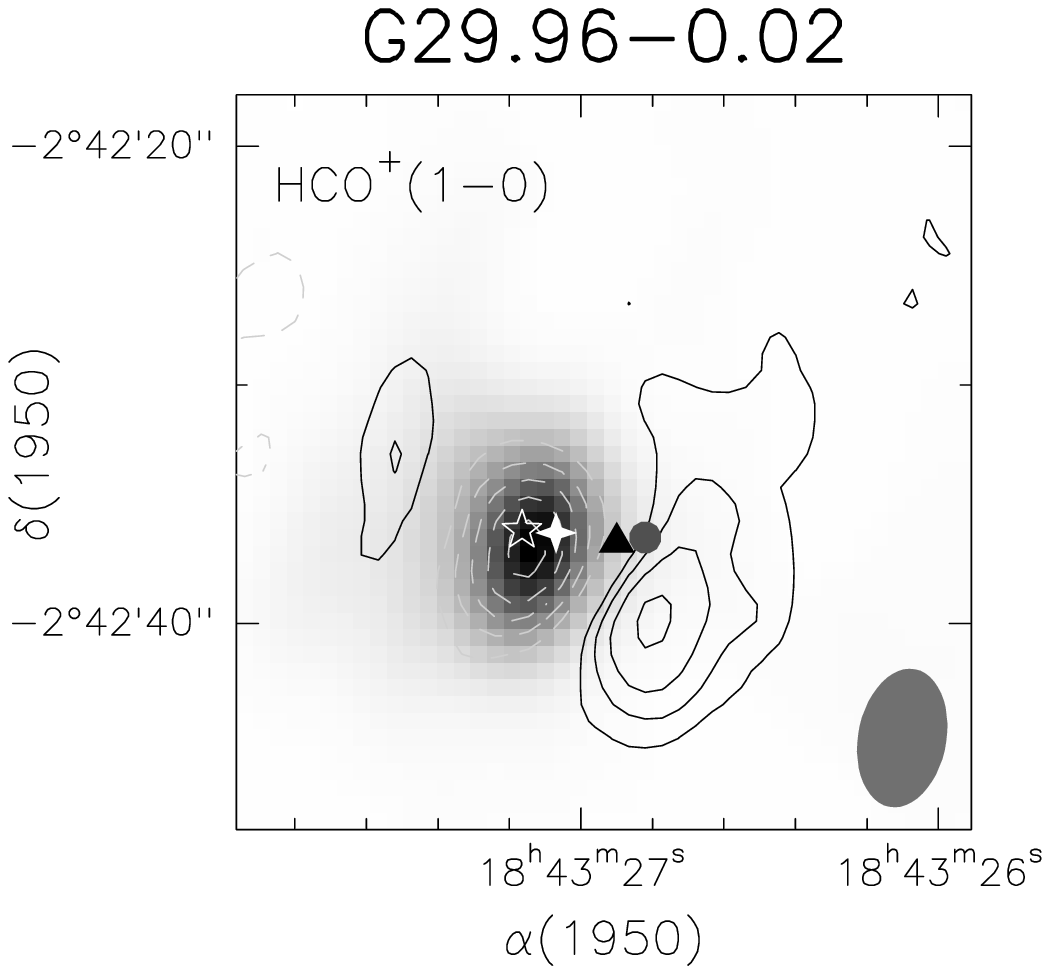,width=6cm}}
\vskip 0.3cm
\centerline{\psfig{figure=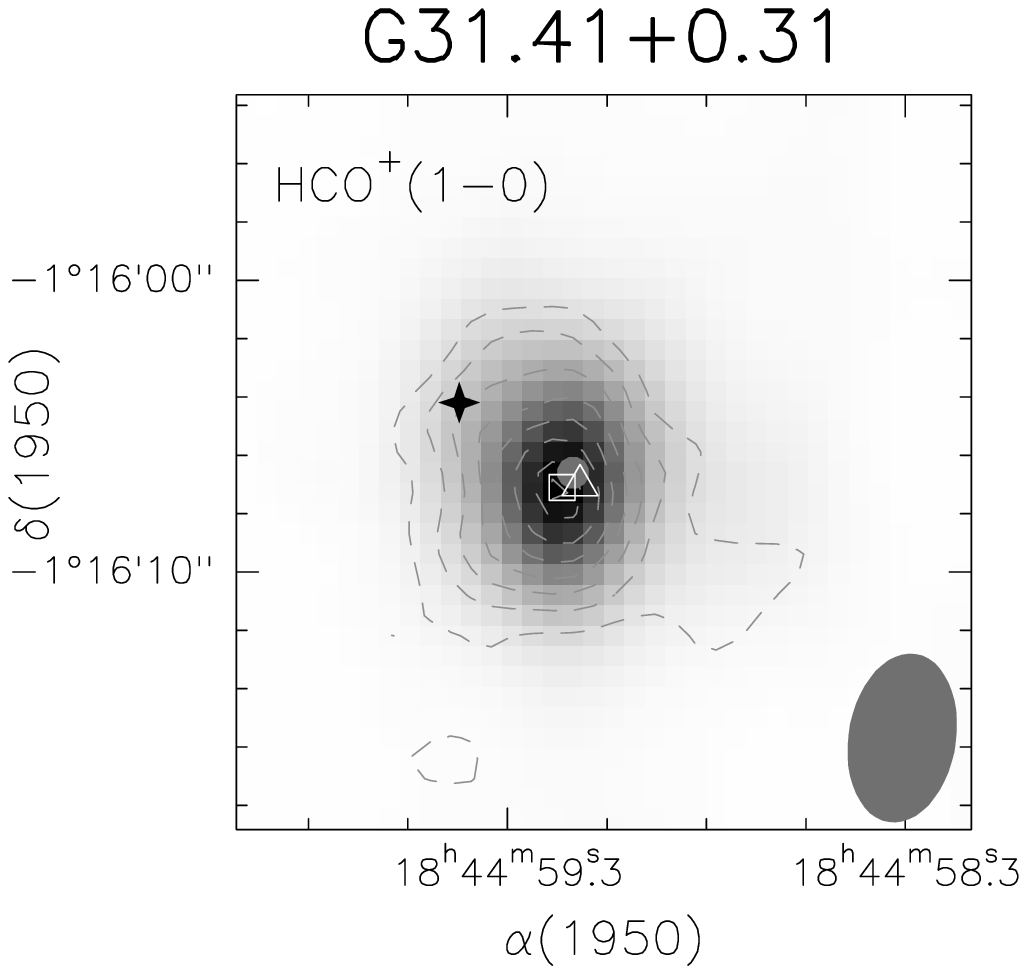,width=6cm}}
\caption{Contour plots of the \hcotr\ line averaged over the
absorption velocity range, 99--102~km/s for \gv\ and 95--99~km/s for \gt,
marked by the dotted lines in Figs.~\ref{g29_spettri} and~\ref{g31_spettri},
overlaid on the 88~GHz continuum emission. Negative contours are dashed.
The contour levels are $-0.5$, $-0.4$, $-0.3$, $-0.2$, $-0.1$,
0.1, 0.15, 0.2,
0.25~Jy/beam for \gv\ and
$-0.42$, $-0.4$, $-0.35$, $-0.3$, $-0.254$, $-0.2$, $-0.15$,
$-0.1$~Jy/beam for \gt.
Symbols are as in Figs.~\ref{g29_righe} and~\ref{g31_righe}.
}\label{hcoass}
\end{figure}

In Figs.~\ref{g29_spettri} and~\ref{g31_spettri} we show the source integrated
spectra for both \hcotr\ and \siotr\ transitions. The \hcotr\ spectra show
absorption at a velocity very close to that of the ambient cloud in the case of \gt, and redshifted
in \gv. No absorption is revealed in the \siotr\ spectra.  From the spectra,
we selected the range of velocity channels highlighted by the vertical bars in
Figs.~\ref{g29_spettri} and~\ref{g31_spettri}, in order to produce maps
of the \hco\ absorption
(dotted bars) and of the line wings (dashed for the red wings and dash-dotted
bars for the blue wings). We produced also maps averaged over all the line emission.

The contour maps of the averaged \hcotr\ and \siotr\ emission,
superimposed on the corresponding continuum in grey scale, are displayed
in Fig.~\ref{g29_righe} for \gv\ and Fig.~\ref{g31_righe} for \gt. The
molecular emission in our \hco\ and \sio\ maps of \gv\ is elongated toward the
position of the \meth\ clump found by Pratap et al.~(\cite{pratap99}).
Particularly, the \sio\ map shows a second peak close to the
estimated position of the \meth\ outflow centre found by
Pratap et al.~(\cite{pratap99}). As expected, in both sources the region of
the \hco\ absorption is coincident with the position of the 88 GHz continuum
peak, as shown in Fig.~\ref{hcoass}: this proves that it is a \emph{true} absorption feature,
not a self-absorption effect.

Towards \gv, the peak of the molecular emission is coincident with the core seen in
the dust continuum and NH$_3$ emission by Cesaroni et al.~(\cite{cesa94}).
We note that in this source most of the 88~GHz continuum is due to
the free-free emission from the UC \HII\ region, which is slightly displaced
from the dust core (see Sect.~\ref{sobs_cont}).
The lack of \hco\ emission towards the UC \HII\ peak
is due to the absorption (see Fig.~\ref{hcoass}).
The red and blue-shifted wing emissions of both \hco\ and \sio\ appear only
slightly separated along a E-W axis, parallel to that defined by the velocity gradient of the
ammonia core (Cesaroni et al.~\cite{cesa98}).

Towards \gt, the peak of the \hco\ and \sio\ clump is shifted southwards
with respect to the position of the core traced by the
dust emission and
high excitation \amm\ and CH$_3$CN transitions
(Cesaroni et al. \cite{cesa98}; Cesaroni et al. \cite{cesa31.94}).
In the case of \hco, this is probably due to the fact that the  core
is coincident with the continuum peak, where the line is in absorption
(see Fig.~\ref{hcoass}). In this case, the red and blue wings are clearly
separated along a NE-SW axis, parallel to the velocity gradient seen in
\amm\ and \methcy.
The \sio\ red wing overlaps with the position  of the dust continuum emission.

In Table~\ref{tabhco} we report the parameters of the \hco\ line emission. For each
source we report the position, velocity, and synthesised beam brightness
temperature ($T_{{\rm SB}}$) for the emission and absorption peaks.
In the same table we also report the observed ($\Theta_{3 \sigma}$)
and deconvolved ($\theta_{3\sigma}$) angular diameter of the \hco\ molecular
clump, computed from the 3$\sigma$ contour, as explained in Sect.~\ref{sres_cont},
and the deconvolved radius ($r_{3\sigma}$).
In Table~\ref{tabsio} we report the parameters for the \sio\ transition:
position, velocity, and synthesised beam brightness temperature
($T_{{\rm SB}}$) for the emission peak. We also give the values of the
integrated line emission ($\int T\, {\rm d}V$) and the peak velocity
($V_{{\rm LSR}}$) 
from gaussian fits to the spectra obtained by averaging in each spectral
channel the line emission
within the contour corresponding to 50\% of the peak in
Figs.~\ref{g29_righe} and~\ref{g31_righe}. The observed and deconvolved
angular size corresponding to these contours
and the deconvolved linear radius are also reported in Table~\ref{tabsio}.

\begin{table*}
\caption[]{Parameters of the \hcotr\ line emission and absorption} \label{tabhco}
\begin{flushleft}
\begin{tabular}{*{9}{c}}
\noalign{\smallskip}
\hline
\noalign{\smallskip}
source   &   &  $\alpha_{1950}$ & $\delta_{1950}$  & $V_{{\rm LSR}}$  & $T_{{\rm SB}}$ & $\Theta_{3\sigma}$ & $\theta_{3\sigma}$ & $r_{3\sigma}$\\
             &                  &                        &                    &(km s$^{-1}$) & (K) &  (\arcsec) &  (\arcsec) & (pc)        \\
\noalign{\smallskip}
\hline
\noalign{\smallskip}
\gv\         &  emission &      18:43:26.87 &    $-$2:42:37.8 &    95.5 &    25.7    &   15.3    & 14.6     & 0.21    \\
             & absorption &     18:43:27.15  &    $-$2:42:37.1  &    98.8 &    $-$6.0    &     &     &     \\
\gt\          &  emission &     18:44:59.15 &     $-$1:16:11.5   &    92.1 &    17.2    &   14.4   & 13.7     & 0.26    \\
             & absorption &     18:44:59.15  &    $-$1:16:07.3  &    97.2 &    $-$5.1    &      &     &     \\
\noalign{\smallskip}
\hline
\end{tabular}
\end{flushleft}
\end{table*}

\begin{table*}
\caption[]{Parameters of the \siotr\ line emission} \label{tabsio}
\begin{flushleft}
\begin{tabular}{*{10}{c}}
\noalign{\smallskip}
\hline
\noalign{\smallskip}
source     &  $\alpha_{1950}$ & $\delta_{1950}$    & $T_{{\rm SB}}$ & $\int T\, {\rm d}V$ & $V_{{\rm LSR}}$ &  $\Delta V_{\rm FWHM}$ & $\Theta_{50\%}$ & $\theta_{50\%}$ & $r_{50\%}$\\
              &                    &                  &         (K)              &  (K km s$^{-1}$)     &  (km s$^{-1})$             & (km s$^{-1})$        &    (\arcsec)  &    (\arcsec)  & (pc)        \\
\noalign{\smallskip}
\hline
\noalign{\smallskip}
\gv\       & 18:43:26.91 &  $-$2:42:39.2   &          8.1             &         39.3           &  97.9                            &  10.5                    &         7.7 & 6.1          & 0.09    \\
\gt\       & 18:44:58.96 &  $-$1:16:10.8    &           6.6            &         39.8           & 95.7                            &  15.3                     &        8.6 & 7.2        & 0.14    \\
\noalign{\smallskip}
\hline
\end{tabular}
\end{flushleft}
\end{table*}

\section{Discussion}
\label{sdisc}

\subsection{Mass estimates}

Assuming optically thin dust thermal emission at 88~GHz, we calculated the
mass of the millimeter continuum sources using:
$M_{\rm d} = \frac{F_\nu d^2}{\kappa_\nu B_\nu(T)}$,
where
$F_\nu$ is the observed integrated flux density on dust emission map
(i.e. after subtracting the computed free-free emission),
%at 88~GHz for \gv, and at 88 and 233.26~GHz for \gt,
$d$ the distance from the Sun,
$B_\nu(T)$ the Planck function at the assumed dust temperature $T$,
$\kappa_\nu=\kappa_{230 \mbox{ \scriptsize GHz}}
{\left(\frac{\nu}{230 \mbox{ \scriptsize GHz}}\right)}^\beta$
is the dust opacity per gram with
$\kappa_{230 \mbox{ \scriptsize GHz}}= 0.005 \mbox{ cm}^2 \mbox{ g}^{-1} $,
assuming a gas-to-dust ratio of 100 by mass (Preibisch et
al.~\cite{preibisch93}), and $\beta$ is taken from Table~\ref{tabfitsed}.

The gas kinetic temperatures of the two HCs have been estimated by
Cesaroni et al.~(\cite{cesa94}) using the \ammtr\ line; they found
83 K for \gv\ and 120 K for \gt. This latter value is consistent
with the estimate of 123~K based on the CH$_3$CN observations by
Cesaroni et al.~(\cite{cesa31.94}).
We used the temperature derived from the hot molecular gas, rather than the
lower temperature derived from the global SED fit, because the latter is an
estimate of the extended envelope temperature. 
In fact, the temperature is determined by the grey-body spectrum peak, 
i.e. the IRAS flux values, wich are sensitive to the extended dust emission,  
while the continuum emission detected in our 88~GHz maps is dominated 
by the hot dust within the HC.
With these temperatures, we estimated the mass of the hot core responsible
for the 88~GHz dust continuum emission: 2900 $M_\odot$ for \gv\
and 1500 $M_\odot$ for \gt.
The latter is consistent with the estimate obtained from the 233~GHz flux for
the same source
(1200~$M_\odot$). While, for \gv\ due to the problems discussed
in Sect.~\ref{sobs_cont},
the 233~GHz map cannot be used to derive a mass estimate.

The total mass of the molecular clumps surrounding the UC \HII\ and HC can be
estimated from the HCO$^+$ and SiO transitions. Here we will
attempt to evaluate the masses using the virial and LTE assumptions.
Given the velocity structures and complex line shapes observed in both
lines, these estimates are affected by large uncertainties.

Assuming that the \hcotr\ is optically thick, the peak $T_{\rm SB}$
(Table~\ref{tabhco}) can be used as an estimate of the gas kinetic temperature.
Another estimate of the temperature of the molecular clump as a whole
can be obtained from the continuum SED fits presented in Sect.~\ref{sres_cont}.
These two estimates are consistent for \gv, while for \gt\ the \hco\ estimate
is much lower.
This discrepancy probably arises from the fact that in this source the
absorption occurs towards the molecular clump peak, while in \gv\ the molecular
clump, centered on the HC, is shifted from the position of the UC \HII\  where
the absorption is centered. We decided to adopt an average value of 30~K
for both clumps.
We assumed an \sio\ abundance of 7$\times$10$^{-11}$ with respect to H$_2$,
following the average value found by Codella et al.~(\cite{CBR99})
for the quiescent SiO component in a sample of young stellar objects.
This value is also in agreement with the abundance derived by McMullin et
al.~(\cite{MM00}) for the quiescent gas in the Serpens core.
Using the integrated line emission given in Table~\ref{tabsio},
we obtained a mass of 1600~$M_\odot$ for \gv\ and 3200~$M_\odot$ for \gt.
The main uncertainties in these estimates are the assumed \sio\ abundance
and LTE-equilibrium condition; in fact the \sio\ molecule abundance
is well known to be enhanced by a factor 10--100 in shocked regions (Bachiller
\& P\'erez Guti\'errez~\cite{BPG97}; Codella et al.~\cite{CBR99}).
Such abundance enhancement would lead to unrealistically high 
mass estimates, unless the SiO is confined to an outflow.

To derive the virial mass from the \siotr\ transition we
used the FWHM calculated from the gaussian fit (Sect.~\ref{sres_lines},
Table~\ref{tabsio}). We estimated a mass of
2500~$M_\odot$ for \gv\ and 8000~$M_\odot$ for \gt.
Owing to the fact that the \sio\ emission
might arise from outflowing gas,
which would increase the line FWHM,
these values should be considered as upper limits.
In the case of \hcotr, the presence of the absorption feature makes very
difficult to determine the intrinsic FWHM of the line.
We fitted the line profile with two gaussian components, one in
emission and one in absorption. The resulting linewidths of the 
emission component
are 6.1 and 7.2~\kms\ for \gv\ and \gt, respectively. The resulting virial mass estimates,
using the sizes from Table~\ref{tabhco}, are 2000 and 3400~$M_\odot$ for
\gv\ and \gt, respectively.
However, the fit is strongly biased by the line wings and thus the 
resulting virial mass is most probably overestimated.
This problem is particularly evident in \gt, where
these features are detected in both \sio\ and \hco, and the virial mass
estimates are less reliable and a few times higher than the masses derived from
the dust emission or using the LTE approximation.

%\begin{table*}
%\caption{Assumed parameters and calculated mass of the sources of dust and line emission.
%The FWHM are calculated from a two-components gaussian fit for the \hco\ spectra, and a gaussian fit for the \sio\ spectra.} \label{tabmassa}
%\begin{flushleft}
%\begin{tabular}{*{10}{c}}
%\noalign{\smallskip}
%\hline
%\noalign{\smallskip}
%Source&d&$T_{\rm d}$&$\beta$&M&$\Delta{\rm V}_{\rm FWHM}$ &  $M_{vir}$ (50\%) & $\Delta{\rm V}_{\rm FWHM}$&  $M_{\rm vir}$ (3$\sigma$) &  $M_{\rm LTE}^{ \rm (a,b)}$ \\
%          &              &                   &                     &     (dust)    &  (\sio) &   (\sio)  & (\hco)& (\hco)&   (\sio)   \\
%         & (Kpc)     & (K)               &                   & ($M_\odot$) &  (km/s) & ($M_\odot$)&  (km/s)  & ($M_\odot$)& ($M_\odot$)\\
%\noalign{\smallskip}
%\hline
%\noalign{\smallskip}
%\gv&6  & 83&2&2900&10.5&2500& 6.1&2000 &1600 \\
%\gt&7.9&120&1&1500&15.3 &8000 &7.2  & 3400 & 3200\\
%\noalign{\smallskip}
%\hline
%\noalign{\smallskip}
%\multicolumn{10}{l}{\scriptsize $^{\rm a }T \sim 30$~K} \\
%\multicolumn{10}{l}{\scriptsize $^{\rm b }N{\rm(SiO)}/ N{\rm(H_2)}\sim 7.\times 10^{-11}$}\\
%
%\end{tabular}
%\end{flushleft}
%\end{table*}

In conclusion, considering all the uncertainties in each of the methods
discussed above, our estimates of the masses of the molecular clumps in
both sources are consistent with 2000~$M_\odot$ with an error of $\sim$50\%.

\subsection{Gas kinematics}

\begin{figure} \centerline{\psfig{figure=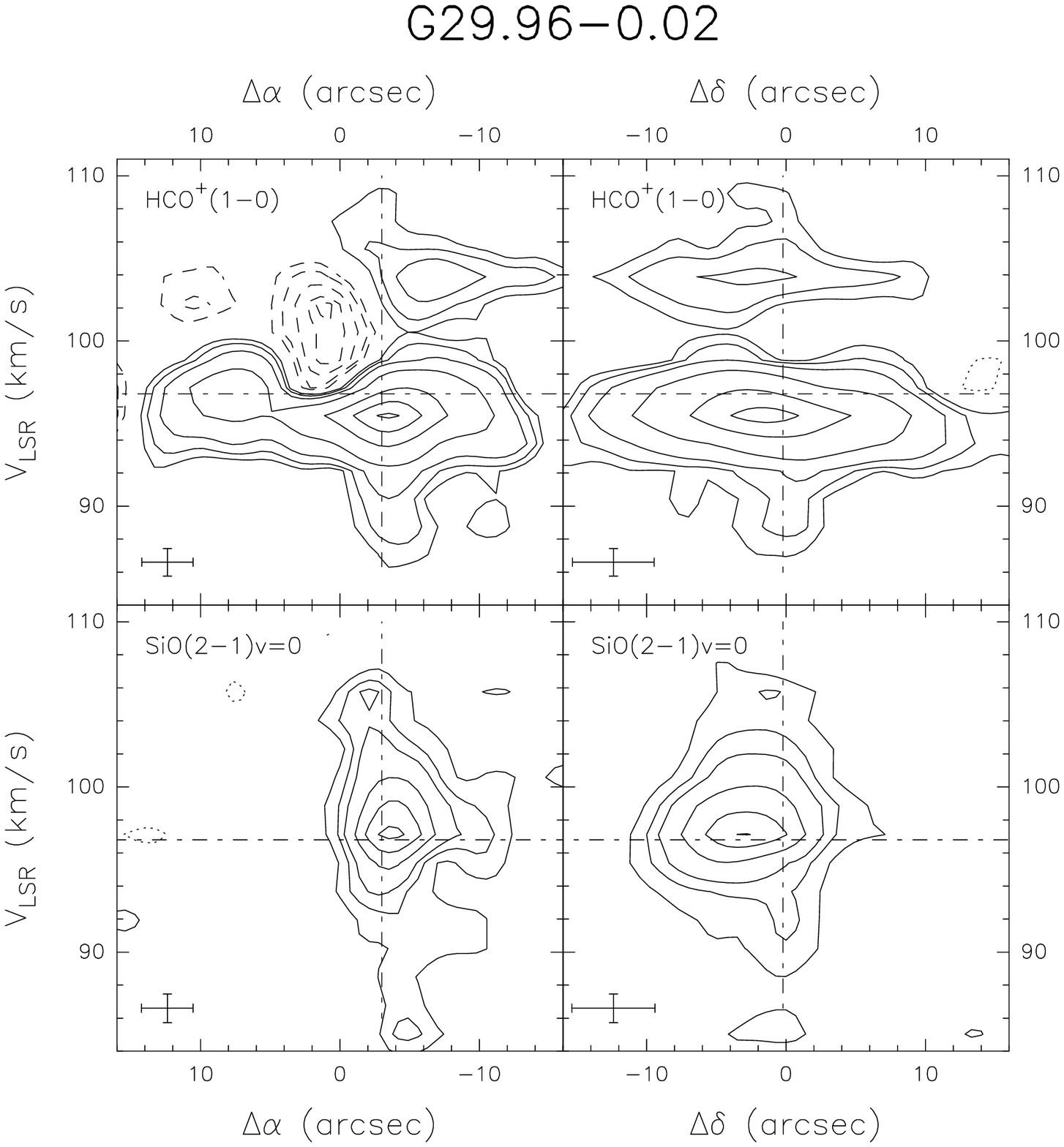,width=8.0cm}}
\caption{Contour plots for \gv\ as a function of $V_{LSR}$ and offset
from the phase center along the E-W axis from the \HII\ region and the
core peak (left) and perpendicularly to it (right).  The maps have been
obtained by averaging the \hcotr\ (top) and \siotr\ (bottom) emission in \gv\
over 8\farcs4 along the E-W axis and over 7\arcsec\ perpendicularly to it.
The contour levels are: for  \hco, in $\Delta\alpha$  $-0.4$, $-0.35$,
$-0.25$, $-0.15$, $-0.09$, 0.09, 0.15, 0.25, 0.50, 1.0, 1.5, 2.0 Jy/beam,
in $\Delta\delta$
$-0.09$, 0.09, 0.15, 0.25, 0.50, 1.0, 1.5 Jy/beam; for \sio,
in $\Delta\alpha$ $-0.15$, $-0.09$, 0.09, 0.15, 0.20, 0.30, 0.40, 0.50 Jy/beam,
in $\Delta\delta$ $-0.09$, 0.09, 0.15, 0.20, 0.40, 0.49 Jy/beam.
The dot-dashed lines mark the position and the velocity of the core detected in
\ammtr\ by Cesaroni et al.~(\cite{cesa98}).}\label{g29_tagli}
\end{figure}

\begin{figure*}
\centerline{\psfig{figure=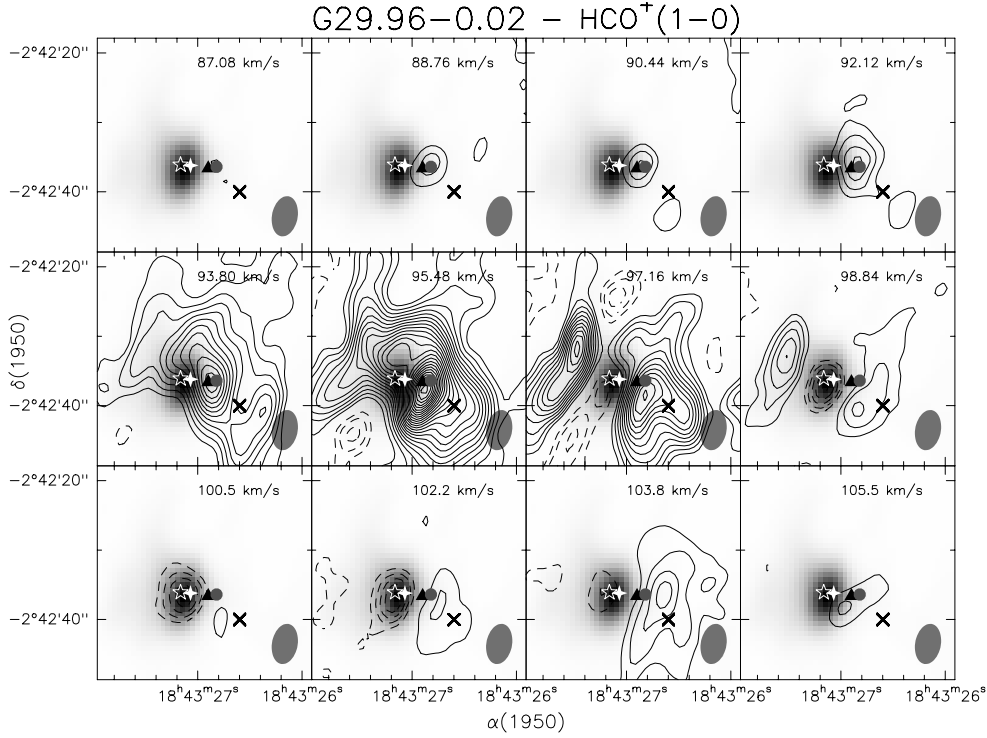,width=13cm,angle=-90}}
\caption{Channel maps of \hcotr\ line towards \gv, overlaid on the 88~GHz
continuum emission. Contour levels are from $-0.55$ to $-0.15$
and from 0.15 to 2.45 by  0.1 Jy/beam.
Negative contours are dashed.  Symbols are as in Fig.~\ref{g29_righe}.
}\label{g29_mapch}
\end{figure*}

We investigated the presence of a velocity gradient along the axes parallel
and perpendicular to the direction where such a gradient was found
in the \ammtr\ (Cesaroni et al.~\cite{cesa98}) and CH$_3$CN(6--5) transitions
(Cesaroni et al.~\cite{cesa31.94}; Olmi et al.~\cite{olmi96}).
For both the \hcotr\ and \siotr\ lines we averaged the emission along the axis
from the UC \HII\ region peak to the core, and perpendicularly to it.  This
axis is along the E-W direction for \gv\ and along the NE-SW direction
for \gt.  In Figs.~\ref{g29_tagli} and~\ref{g31_tagli} the
position-velocity diagrams are shown, respectively for \gv\ and \gt: we
plotted, in the top panles for  \hcotr\ and in the bottom
for \siotr, the offset along the axes previously described.
In \sio\ a velocity-position trend
is clearly seen along the ``parallel'' axes.  The $V_{\rm LSR}$ varies
from 89 to 107~km~s$^{-1}$ over $\sim$3\arcsec\ toward \gv\ and from 86
to 106~km~s$^{-1}$ over $\sim$4\arcsec\ toward \gt.  The corresponding
velocity gradients are approximately 6~km~s$^{-1}$ arcsec$^{-1}$ toward
\gv\ and 5 km~s$^{-1}$ arcsec$^{-1}$ toward \gt, consistent with the
velocity gradient of approximately $2-3$~km~s$^{-1}$~arcsec$^{-1}$ over
3\arcsec\ found along the same directions in ammonia by Cesaroni et
al.~(\cite{cesa98}).  A velocity gradient of the same order is seen in
\hco\ toward \gt, while it is not evident in the case of \gv, most likely
because of the presence of absorption. In the following we propose an
interpretation of these findings

\subsubsection{\gv}

\begin{figure}
\centerline{\psfig{figure=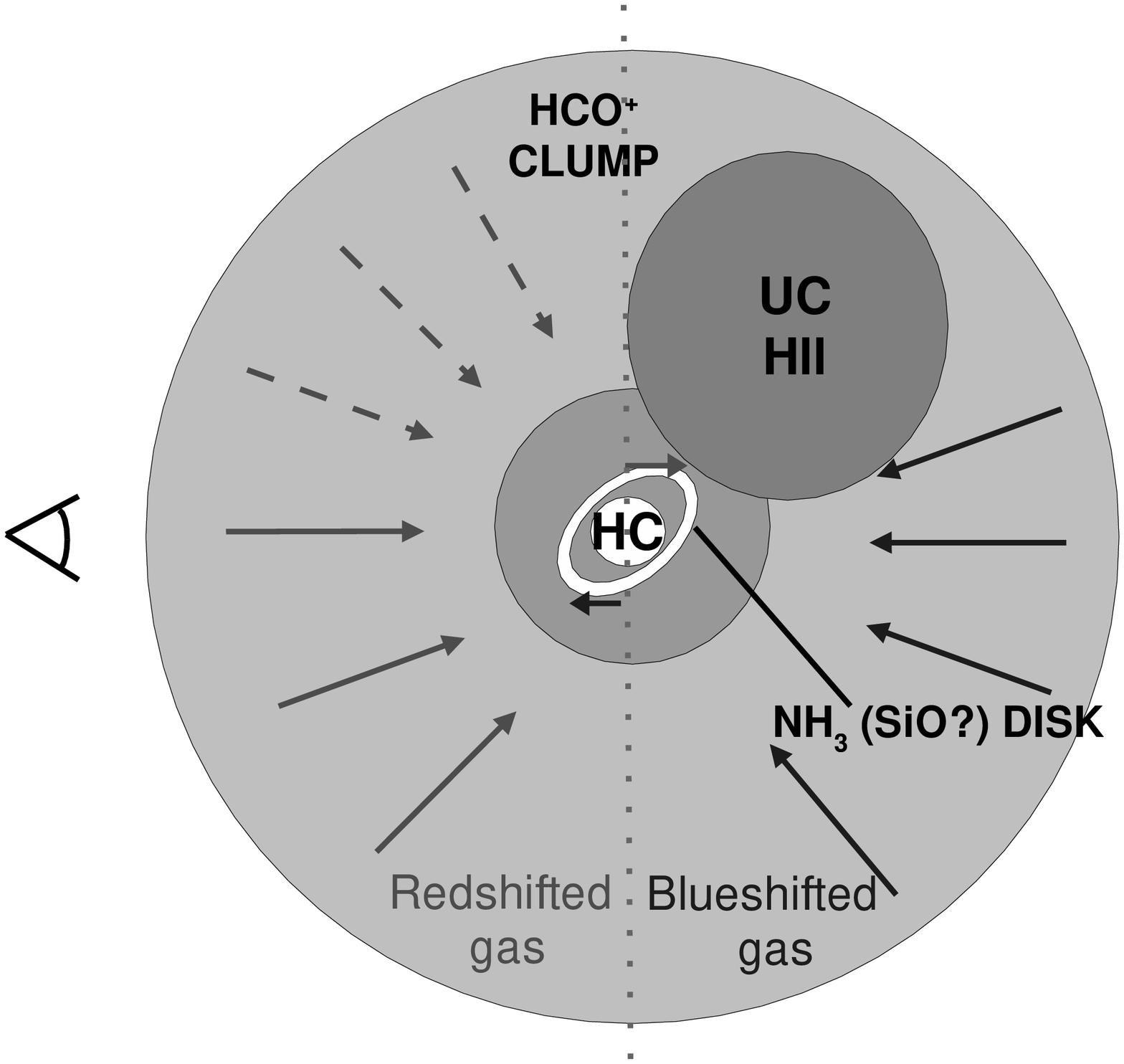,width=8cm}}
\centerline{\psfig{figure=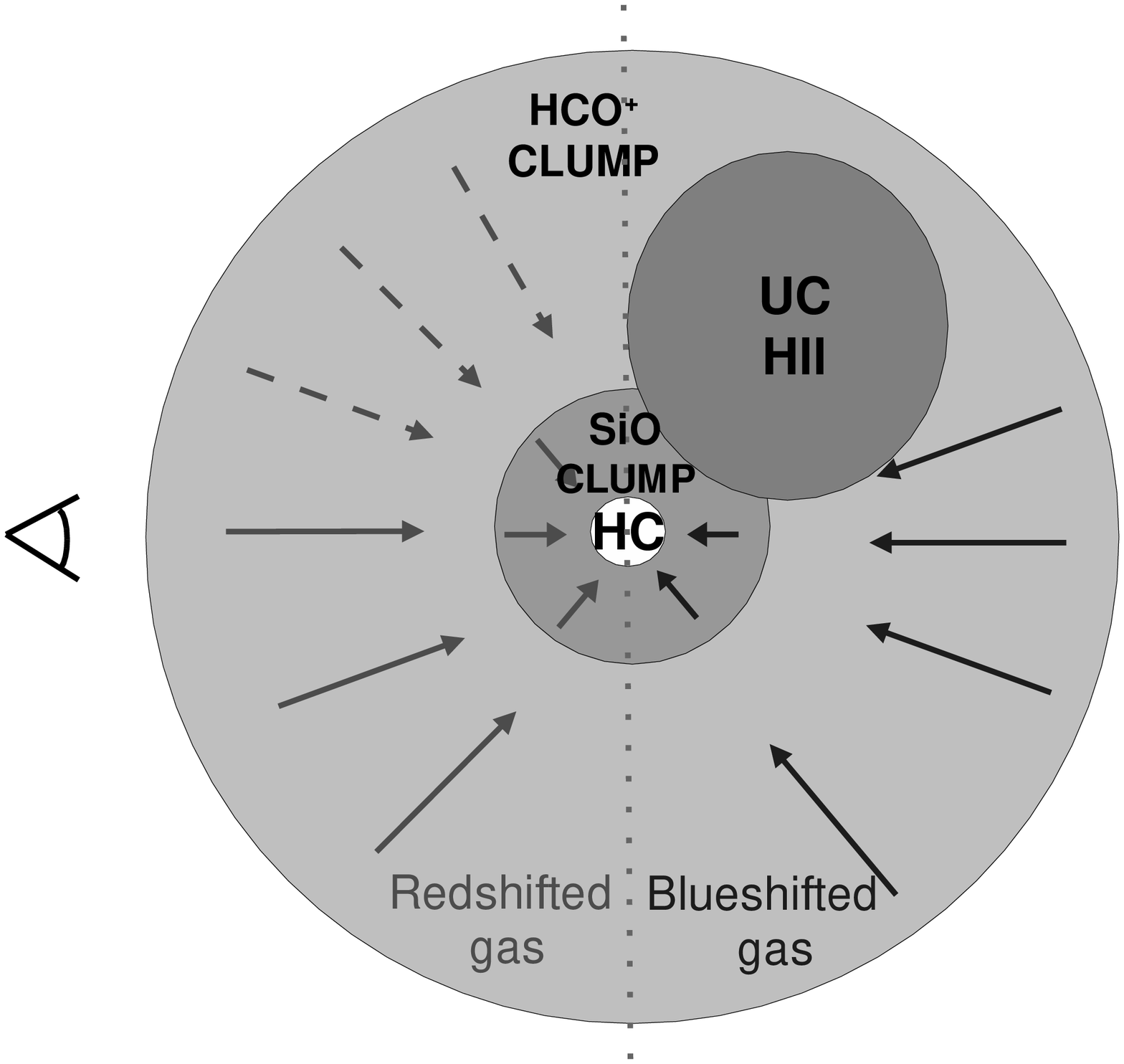,width=8cm}}
\caption{Sketch of the infall model for \gv.
The \sio\ could trace the outer regions of the rotating \ammtr\ disk (top) or
the inner regions of the infalling molecular clump (bottom).
Arrows show the gas velocity with respect to the clump $V_{LSR}$. 
The dashed arrows represent the gas seen in absorption toward the
UC~\HII. The observer is on the left side.
}\label{g29_mod}
\end{figure}

In the case of the \gv\ molecular clump,
as shown in the position-velocity diagrams (Fig.~\ref{g29_tagli}),
the center of the \hco\ clump is coincident in both position and velocity
with the \amm\ HC; which in turn is located
at the geometrical center of the molecular clump, as shown in Fig.~1 of
Kurtz et al.~(\cite{kurtz00}).
The \hco\ absorption appears redshifted with respect to the molecular
clump systemic $V_{\rm LSR}$ (as traced by C$^{34}$S and C$^{17}$O; Cesaroni
et al.~\cite{Cea91}, Hofner et al.~\cite{Hea00}),
and positionally coincident with the UC \HII\ region, located east of the emission
geometrical center. \hco\ channel maps in Fig.~\ref{g29_mapch} also
show blueshifted gas emission toward the same direction, suggesting that
the gas behind the \HII\ region is moving towards  the observer.
This implies that the absorbing gas is moving along the
line of sight towards the UC \HII\ region. Both pv-diagrams also show that
the maximum line-of-sight gas velocities are in the direction of the core
position.
%Finally, the red and blue wing emission slight separation
%(Fig.~\ref{g29_righe})
%is due to the \HII\ region, which absorbs the
%blueshifted emission from the gas behind the HC (to the left of the core).

All these features can be easily interpreted if the UC \HII\ is embedded
within an infalling, spherically symmetric \hco\ clump centered on the
HC position.
As shown in the sketch in Fig.~\ref{g29_mod},
if the clump is contracting toward the \amm\ HC, toward the UC~\HII\ we will
observe redshifted absorption from the gas in front of it
and blueshifted emission from the
gas behind it. At the position of the HC, there is emission at
the highest line of sight velocities for both the redshifted  and
blueshifted gas.

The less abundant, optically thin \sio\ molecule traces a more internal
region of the molecular clump. For this molecule, the pv-diagrams
show a peak coincident with the HC, within the uncertainties, and a
velocity structure consistent with that derived from the \ammtr\ observations of
Cesaroni et
al.~(\cite{cesa98}). To explain this feature,
we can hypothesize that
\sio\ could trace the outer regions of the rotating \ammtr\ disk
or the inner regions of the infalling molecular clump (see Fig.~\ref{g29_mod}).
In the first scenario, the velocity gradient would be due to the velocity
field in the outer regions of the disk. In this view the \sio\ abundance
could be enhanced at the interface between the infalling molecular
gas and the surface of the disk.
Altenatively, the \sio\ emission could trace the innermost and denser regions
of the infalling envelope: the observed velocity gradient
could then be produced by the asymmetric erosion by the UC~\HII. If the
ionised gas is sligthly off-center and further away from the observer
with respect to the center of the infalling clump (as sketched in
Fig.~\ref{g29_mod}), blue-shifted \sio\ gas is eroded from one of the
sides of the clump giving the illusion of a velocity gradient accross the core.
It should be noted,
however, that the \sio\ is generally thought to trace outflowing 
material, as shown by several authors (Bachiller \&
P\'erez~Guti\'errez~\cite{BPG97}; Cesaroni et al.~\cite{Cea99}; Acord et al.~\cite{Aea97}), therefore one cannot exclude {\it a priori} the possibility
that the velocity structure  observed at the HC position
(Fig.~\ref{g29_righe}, bottom panels) may be due to a compact jet-outflow
component unresolved by our observations.

At the position of the \meth\ core detected by Pratap et al.~(\cite{pratap99})
we detect emission in both \sio\ and \hco. We do not detect any clear
signature of outflowing gas at that location, and favour the interpretation
of a second, less massive molecular clump. Future higher resolution observations
may allow to clarify the matter definitively.

\subsubsection{\gt}

\begin{figure}
\centerline{\psfig{figure=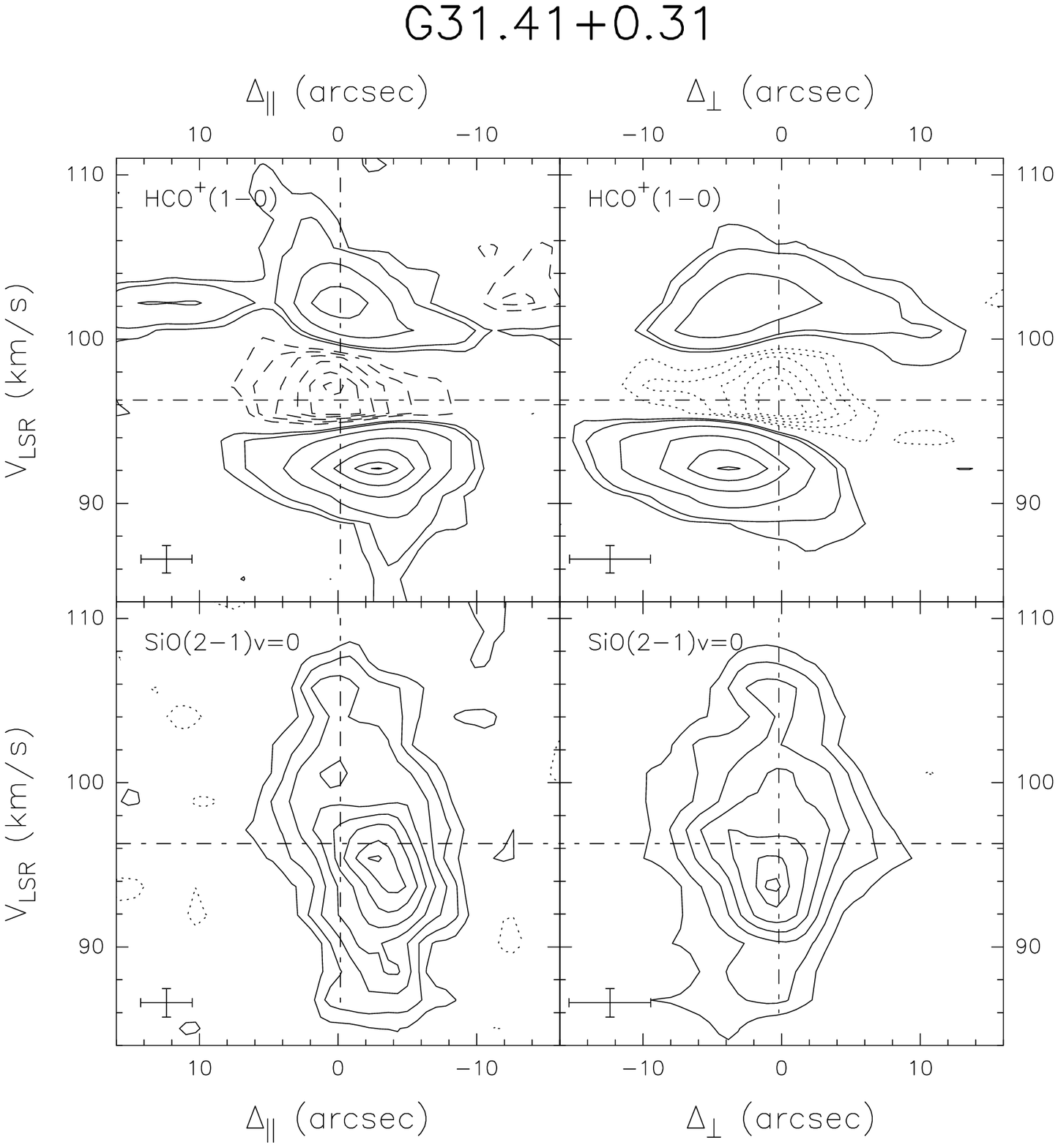,width=8.0cm}}
\caption{Contour plots for \gt\ as a function of $V_{LSR}$ and offset from the
phase center along the SW-NE axis from the \HII\ region and the core peak
(left) and perpendicularly to it (right). The maps have been obtained by
averaging the \hcotr\ (top) and \siotr\ (bottom) emission in \gt\  over
10\farcs5 along the SW-NE axis and  perpendicularly to it.  The contour levels
are: for  \hco, in $\Delta_{\parallel}$ $-0.30$, $-0.25$, $-0.20$, $-0.15$,
$-0.10$, $-0.06$, 0.06, 0.10, 0.25, 0.50, 0.75, 1.0, 1.12 Jy/beam,
in $\Delta_{\perp}$
$-0.3$, $-0.25$, $-0.20$, $-0.15$, $-0.10$, $-0.06$,
0.06, 0.10, 0.25, 0.50, 0.75, 1.0 Jy/beam;
for \sio, in $\Delta_{\parallel}$ $-0.1$, $-0.06$, 0.06, 0.10, 0.15,
0.20, 0.25, 0.30, 0.35, 0.38 Jy/beam, in $\Delta_{\perp}$ $-0.09$, $-0.06$,
0.06, 0.10, 0.15, 0.20, 0.25, 0.30, 0.32 Jy/beam.
The dot-dashed lines mark the position
and the velocity of the core detected in \ammtr\ by Cesaroni et
al.~(\cite{cesa98}).  }\label{g31_tagli}
\end{figure}

\begin{figure*}
\centerline{\psfig{figure=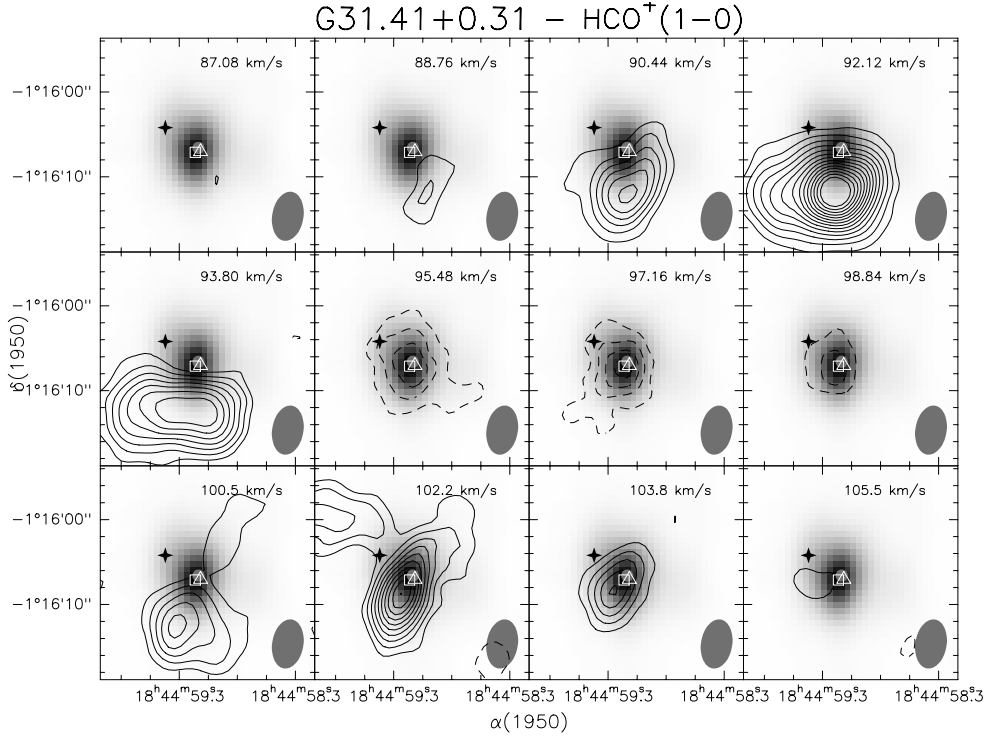,width=13cm,angle=-90}}
\caption{As fig.~\ref{g29_mapch} for \gt.  The contour levels are
from $-0.45$ to $-0.15$ and from 0.15 to 1.65 by  0.1 Jy/beam.
Symbols are as in Fig.~\ref{g31_righe} }\label{g31_mapch}
\end{figure*}

Compared to \gv, the structure of \gt\ is more difficult to interpret.
In this case,
the bulk of the free-free emission from the UC~\HII\ region is
along the same line of sight of the \ammtr\ HC detected by
Cesaroni et al.~(\cite{cesa94}, \cite{cesa98}).
Due to the absorption at the position of the HC,
the observed molecular line emission peak, both in \hco\ and \sio,
is south of the HC (Figs.~\ref{g31_righe} and~\ref{hcoass}).

The pv-diagrams (Fig.~\ref{g31_tagli}) show at high velocity,
$V \ge 8~{\rm km/s}$ from the HC $V_{\rm LSR}$, in both transitions, a gradient
consistent with that observed in \ammtr\ (Cesaroni et al.~\cite{cesa98}).
At lower velocity, however, the situation in \hco\ is more complicated,
with emission and absorption occurring toward the same line of sight
at both blueshifted and redshifted velocities, with respect to the $V_{\rm LSR}$ of the
HC and of the molecular clump as seen in C$^{34}$S and C$^{17}$O
(Cesaroni et al.~\cite{Cea91}; Hofner et al.~\cite{Hea00}).
The channel maps (Fig.~\ref{g31_mapch}) clearly show the
superposition of these velocity features, and show an extension of the emission
at blue velocities to the south of the HC position.
In \sio, the emission peak lies south-west of the \amm\ HC and is blueshifted with respect to
the latter (Figs.~\ref{g31_righe} and~\ref{g31_tagli}).
The red emission (Fig.~\ref{g31_righe}) peaks toward the HC,
whereas the line bulk and the blue emission peak south-west of the HC,
the velocity gradient agreeing with that seen in \amm.

It seems that
there is a north-south gradient from red to blueshifted emission in
\hco, possibly tracing an outflow or the outer regions of the \amm\ disk,
and a NE-SW gradient in \sio, compatible with the disk hypothesis. Thus,
while the observed velocity field could also be interpreted
as a combination of infall, outflow and rotation, as it may happen in similar
regions (see e.g. Wilner et al.~\cite{Wea96}), it is clear that we lack
the spatial resolution required to properly separate the various components.
In fact, at present we cannot exclude the possible
alternative explication that we are actually observing the superposition
of two molecular cores, one located at the \amm\ HC position and the second
south of it, where the \hco\ and \sio\ emission peak. In this view, the
velocity structure we observe would be partially produced by the superposition of the
velocities of the two cores.
However, we note that unlike \gv, in \gt\ the absorption dip is seen at the centre of the
\hco\ line: this seems to be more consistent with the disk hypothesis, as 
collapse
and outflow should determine a shift in velocity of such an absorption dip.
In conclusion, we would 
tend to favour the disk hypothesis, although possibly in a
more complex scenario involving outflow and/or infall.

\section{Conclusions}
\label{sconc}

Our interferometric observations show that the HCs and UC~\HII\ in
\gv\ and \gt\ are embedded within massive gaseous clumps with a complex
velocity structure. The masses of the clumps as derived from 
the millimeter continuum emission and from virial
and LTE estimates based on the molecular line emission
are in the range 1000--3000~$M_\odot$ for both sources (with \gt\ being 
sligthly more massive than \gv). 
In \gv\ the HC is located close to the geometrical center of the 
clump and there is compelling evidence for a global motion of the
molecular material onto the central object. In \gt\ our spatial resolution 
is not high enough to resolve the various components and it is not
easy to disentangle the observed velocity structure. Most likely,
in this source several kinematical components (outflow, rotation and possibly
infall) are merged together by the spatial resolution of our observations.

In both regions our observations are consistent with earlier claims
of the presence of rotating disks around young accreting massive (proto-)stars
(Cesaroni et al.~\cite{cesa98}). However, our spatial resolution does not 
allow us to definitely prove such claims and alternative models to interpret
our observations are still viable.
To definitely solve this dilemma and to probe the detailed structure 
of \gt, higher resolution observations are needed.

\begin{acknowledgements}
The Owens Valley millimeter-wave array is supported by NSF grant
AST-96-13717. Research at Owens Valley on the formation of young
stars and planets is also supported by the {\it Norris Planetary Origins
Project}. This work was
partly supported by  ASI grant  ARS-98-116
to the Osservatorio di Arcetri.
\end{acknowledgements}

\end{document}